\documentclass[aps, amsmath,amssymb,prb,twocolumn,superscriptaddress]{revtex4-2}
\usepackage{graphicx}
\usepackage[colorlinks,bookmarks=false,citecolor=blue,linkcolor=red,urlcolor=blue]{hyperref}
\usepackage[utf8]{inputenc}
\usepackage{amsmath,bm}
\usepackage{bbold}
\usepackage{color,soul}
\usepackage{xstring}
\usepackage{wasysym}
\usepackage{xspace}
\usepackage{amssymb}
\usepackage{time}
\usepackage{bbold}
\usepackage{subfigure}
\usepackage{multirow}
\usepackage{xspace}
\usepackage{url}
\usepackage{amsmath}
\usepackage{booktabs}
\usepackage{lipsum}

\def\tg{\tilde{\Gamma}}
\def\<{\langle}
\def\>{\rangle}
\DeclareMathOperator{\Tr}{Tr}

\usepackage{ulem}

\begin{document}

\title{
Probing the critical behavior of a sign-problematic model with Monte Carlo simulations}
\author{\firstname{Ye} \surname{Ling}}
\email{These authors contributed equally to this work.}
\affiliation{\mbox{School of Physical Science and Technology, Beijing University of Posts and Telecommunications, Beijing 100876, China}}
\affiliation{\mbox{School of Physics and Astronomy, Beijing Normal University, Beijing 100875, China }}

\author{\firstname{Yuting} \surname{Wang}}
\email{These authors contributed equally to this work.}
\affiliation{\mbox{School of Physical Science and Technology, Beijing University of Posts and Telecommunications, Beijing 100876, China}}

\author{\firstname{Wenan} \surname{Guo}}
\email{waguo@bnu.edu.cn}
\affiliation{\mbox{School of Physics and Astronomy, Beijing Normal University, Beijing 100875, China }}
\affiliation{\mbox{Key Laboratory of Multiscale Spin Physics (Ministry of Education), Beijing Normal University, Beijing 100875, China}}

\author{\firstname{Yuhai} \surname{Liu}}
\email{yuhailiu@bupt.edu.cn}
\affiliation{\mbox{School of Physical Science and Technology, Beijing University of Posts and Telecommunications, Beijing 100876, China}}
\date{\today}
\begin{abstract}
The sign-problematic generalized Baxter-Wu (GBW) model with asymmetric complex couplings
%for which the critical properties are exactly known,
is mapped onto a one-dimensional quantum model.
%Through Monte Carlo simulation,
Utilizing the model's exactly known critical properties, we study the relation between the conventional and the modified average signs and the phase transitions in the GBW model.
We find that the average sign develops a negative peak near the critical point, but it is not a unique indicator
of phase transition, as similar features can appear in noncritical regions.
While the average modified sign provides
a viable probe for the phase transition, the practical effectiveness of this method is limited by the exponential scaling of computational cost with the system's volume.
We propose that the universal properties of the original model can be
investigated through simulating the related reference model, based on the universality assumption. Using finite-size scaling analysis based on Monte Carlo simulations, we confirm the validity of this method, which thereby provides a novel framework for investigating phase transitions in systems plagued by the sign problem.

\end{abstract}
\maketitle
\section{Introduction}

 The sign problem poses a significant challenge in the application of Monte Carlo (MC) methods
 across various fields, particularly in quantum Monte Carlo simulations of fermionic or
 frustrated bosonic systems. This issue arises when sampling weights in the MC process
 vary between positive,
 negative, or even complex values. For a system with a sign problem, one can construct an MC importance sampling
 process using the absolute values of the weights, which defines a reference ${\cal Z}_+$ model, and calculate the observables based on a reweighting
 technique\cite{Troyer2005,10.21468/SciPostPhysCodeb.1-v2.4}. However, in such a procedure,
to achieve a given accuracy, the number of MC steps scales exponentially with the system size,
making it impossible to achieve physically meaningful precision within a reasonable timeframe.

Despite extensive efforts made in the past few decades to resolve \cite{Zi-Xiang2015} or alleviate \cite{chang2023boosting, Karakuzu2023, Stefan2017, XiangT2016} the sign problem, it still persists without a universally applicable solution.
In a recent study, Mondaini et al. go the opposite way: rather than striving to eliminate or alleviate the sign problem, they show that the sign problem in determinant quantum Monte Carlo (QMC) is quantitatively linked to quantum critical behavior\cite{Mondaini2022}.
%the relationship between the average sign and phase transitions in quantum many-body systems \cite{Mondaini2022}.
This study charts a path for exploiting the average sign in QMC simulations to understand quantum critical behavior and motivates further exploration of the average sign in various Monte Carlo methods along this direction \cite{Mondaini2023, TianxingMa2023, Manvsen2024}.
Ma et al. \cite{Manvsen2024} point out that the average sign is related to the model ${\cal Z}$ itself and the reference model ${\cal Z}_+ $, the sign can exactly probe the phase transition only if the free energy in the reference system is flat across a range of parameters near the critical point of the original model.
To broaden the application scope of the average sign method, they propose a modified sign that fixes the temperature of the reference system to
eliminate its influence and successfully apply it to detect the critical point of the frustrated $J_1$-$J_2$ easy-axis XXZ model.

To draw a more concrete conclusion on the relationship between the sign problem and phase transitions, a model with a sign problem and exactly known transition points
and critical properties is ideal. The generalized Baxter-Wu (GBW) model is one such candidate. The proper Baxter-Wu model,
which is a statistical physics model describing three spin interactions on a triangular lattice, was
introduced by Wood and Griffiths ~\cite{D_W_Wood_1972} and exactly solved by  Baxter and Wu \cite{Baxter1973}.
It exhibits the critical behavior of the four-state Potts fixed point in two dimensions (2D) without logarithmic corrections
\cite{Baxter_1973, Wu1982, Domany_BW-Potts}.
The model is generalized by allowing for different couplings in the up- and down-triangles, with a phase transition
occurring along a line of self-dual points \cite{Youjin2010, blote2017}.
In the case the couplings are real,  the transition moves away from the four-state Potts fixed point into the first-order region\cite{Youjin2010}.
In contrast, complex couplings bring the transition in the opposite direction, resulting in behavior characterized by
the 2D four-state Potts universality class with logarithmic corrections, as present in the 2D four-state Potts model\cite{blote2017}.
Here, we focus on the case of the complex coupling GBW model. We will show that the model is
equivalent to a one-dimensional quantum model; therefore, the phase transition is related to a quantum phase transition at zero temperature.
By binding two configurations that are related under a $\pi$ rotation about a lattice direction,  we can eliminate the imaginary parts of the corresponding joint weights.
However, these weights still contain cosine factors, ultimately leading to sign problems.
Since the critical point is known via the self-dual relationship, the model
provides an excellent platform for studying the relationship between the criticality and the usual average sign or the modified average sign introduced in Ref.~\cite{Manvsen2024}.

Our analyses reveal that both average sign methods face fundamental limitations. The former shows a nonunique negative peak near criticality, which can also arise in noncritical regions, whereas the latter, though a valid phase-transition probe, suffers from computational costs that scale exponentially with system size.

Furthermore, the reference ${\cal Z}_+$ model-defined by using the absolute values of the weights of binding configurations-shares the same groundstate
degeneracy with the original model. Consequently, the phase transitions in both models belong to the same universality class and exhibit similar finite-size
scaling behavior. In this paper, we will demonstrate that studying the reference model provides a new approach for investigating systems with the sign problem via MC simulations.

%\Yuhai{Collectively,
%our large-scale numerical simulations
In this paper, we demonstrate that while sign-based diagnostics are unreliable or impractical in general, the use of universality considerations provides a viable means of learning critical properties via the reference model of a sign-problematic model.

The paper is organized as follows.  In Sec.~\ref{sec:model},  we describe the GBW model with asymmetric complex coupling and map it to a 1D quantum model.
The model and its sign problem are expressed using the weight of bound configurations. The reference model is also defined and utilized for MC simulations.
In Sec.~\ref{sec:results},  we discuss our MC simulation results. In Sec.~\ref{sec:conclusions} we give our conclusions and discussion.
\section{Model and method}
\label{sec:model}
\begin{figure}[htb]
\includegraphics[width=0.4\textwidth]{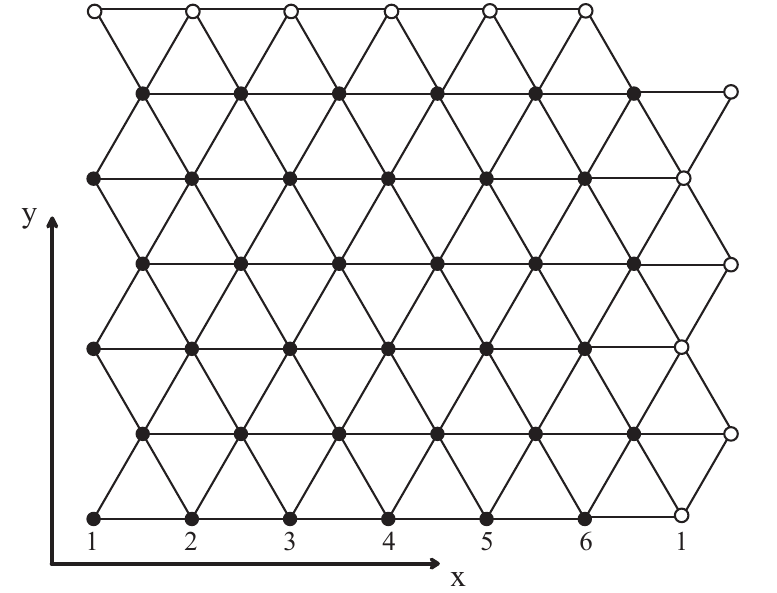}
  \caption{\label{fig:TRI_lattice}Triangular lattice with periodic boundaries.  }
\end{figure}

We consider the generalized Baxter-Wu model with different complex couplings in the up- and down triangles, see Fig.~\ref{fig:TRI_lattice},
with the reduced Hamiltonian
\begin{equation}
\label{Eq:Ham}
H/T=- K_{\rm up}\sum_\triangle \sigma _{i}\sigma _{j}\sigma_{k}-   K_{\rm down} \sum_{\bigtriangledown } \sigma _{l}\sigma _{m}\sigma_{n},
\end{equation}
where spin $\sigma_i$ takes the values $\pm 1$ at site $i$, $T$ is the temperature, and the sums are on the up- and down-triangles of the triangular lattice, respectively, with
$K_{\rm up}=K+\Delta K, K_{\rm down}=K-\Delta K$, and $\Delta K=i \phi$\cite{blote2017}.
The model reduces to the proper Baxter-Wu model when $\Delta K=0$.

\begin{figure}[htb]
 \includegraphics[width=0.5\textwidth]{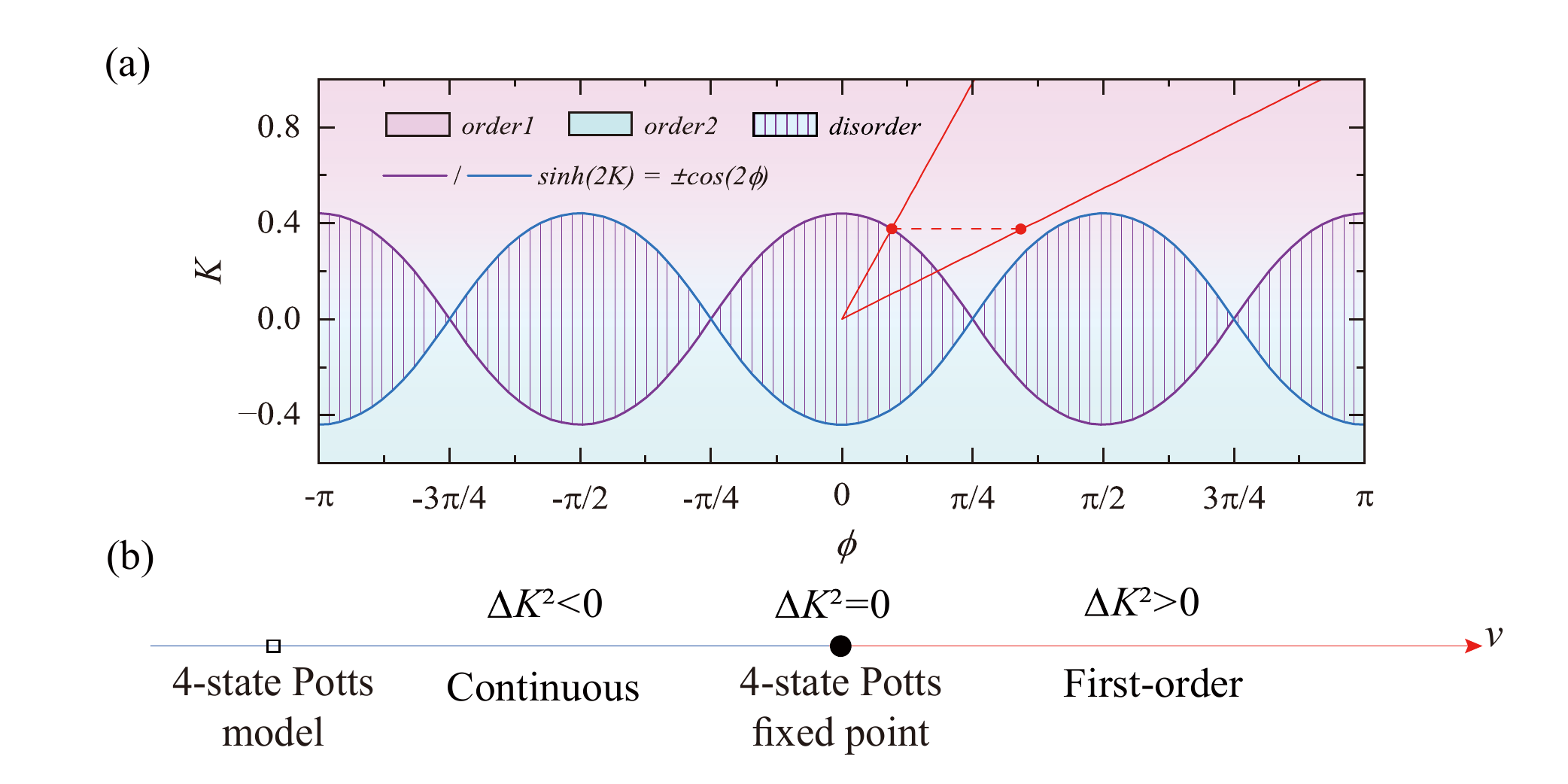}

  \caption{\label{fig:phase_diagram}(a)The phase diagram of the generalized Baxter-Wu model with
  complex couplings. The critical lines dividing ordered and disordered phases are exactly known; The ordered phase ``order1" is ferromagnetic, and another ordered phase ``order2" is antiferromagnetic.
	The disorder phase exists between the two critical lines. The two red lines represent
    the line $K=K_c/T, \phi=\phi_c/T$ going through the critical point ($K_c=0.376165, \phi_c=0.3$) and the line $K=K_c/T, \phi=(0.3+\pi/4)/T$ going through a shifted point $(K_c=0.376165, \phi=0.3+\pi/4)$, respectively.
    (b)The RG flow diagram of the generalized Baxter-Wu model. $\Delta K^2<0$($\Delta K=i\phi$), corresponding to complex couplings, brings the model logarithmic corrections; $\Delta K^2>0$, corresponding to the real couplings, moves the model into the first-order range.}
\end{figure}

\begin{figure}[htb]
  \includegraphics[width=0.45\textwidth]{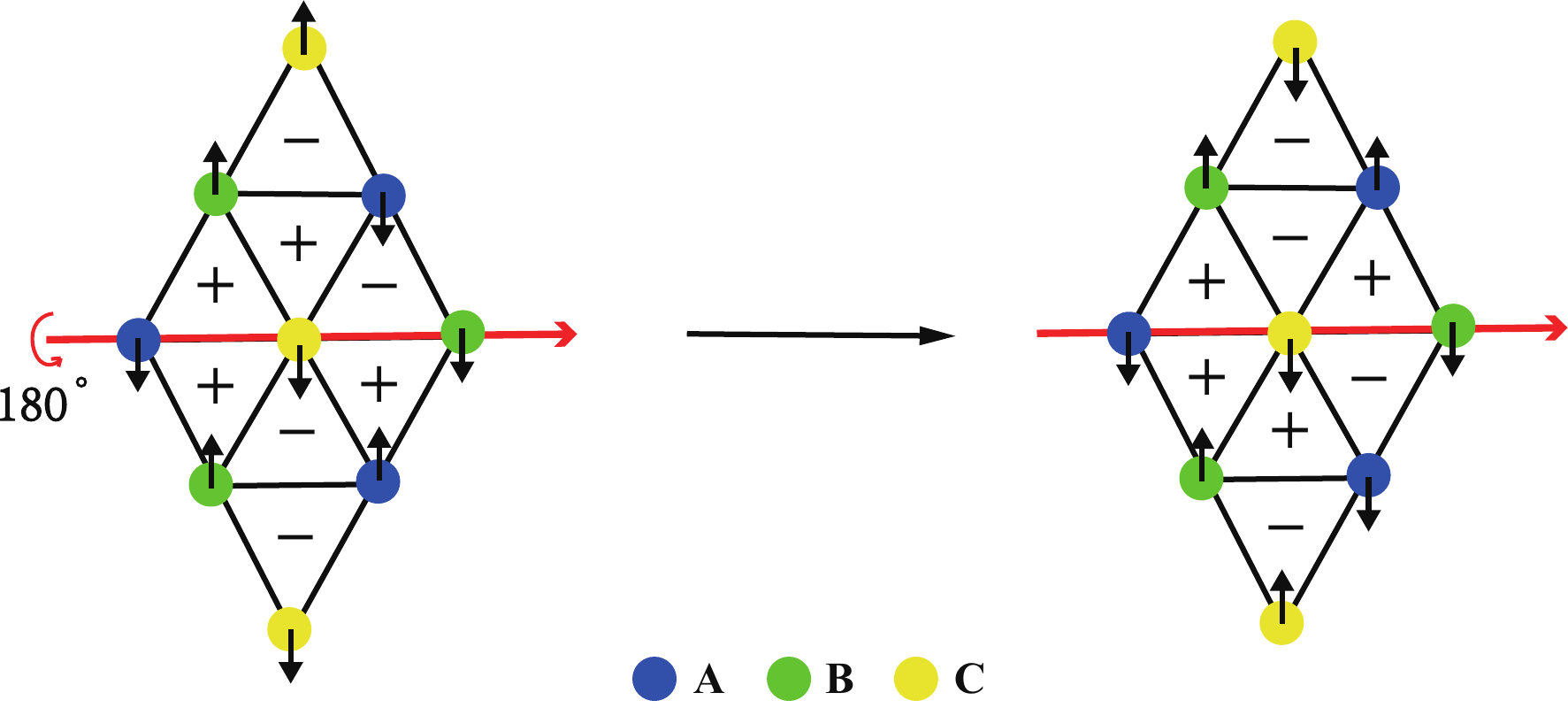}\caption{\label{fig:Binding_configuration}
  A $\pi$ rotation about a lattice direction transforms the system configuration $\Gamma$(left)  into a new configuration $\Gamma'$ (right).
  The up-triangle and down-triangle in the two configurations are exactly reversed. This relationship allows us to bundle the configurations together and eliminate the imaginary parts in their joint weight.}
\end{figure}

The model is self-dual on lines \cite{Youjin2010}
\begin{equation}
	\sinh (2K_{\rm up}) \sinh (2K_{\rm down}) =1,
\end{equation}
which are two critical lines
\begin{equation}
    \sinh(2K)=\pm \cos(2\phi),
    \label{cp}
\end{equation}
separating a disordered phase from two ordered phases: one (order1) is ferromagnetically ordered
for $K>0$, with four ground states satisfying $(\sigma_A, \sigma_B, \sigma_C)=(+,+,+), (+,-,-), (-,+,-), (-,-,+)$ on every triangle which contains sites of sublattices $A, B, C$; the other one is antiferromagnetically ordered for $K<0$, with four ground states obtained by flipping all spins of the ferromagnetic ground states.
The phase diagram is shown in Fig.~\ref{fig:phase_diagram}(a).

The proper Baxter–Wu model ($K_{\rm up}=K_{\rm down})$ displays the critical behavior of the 2D four-state Potts model\cite{Domany_BW-Potts}.
In particular, the model sits precisely at the $q=4$ Potts fixed point, apart from other irrelevant fields; therefore, the logarithmic corrections seen in the
four-state Potts model due to a marginal dilution field $v$ absent.
The introduction of different couplings in the up- and down triangles moves the model away from this fixed point by making $v \propto \Delta K^2$ \cite{blote2017}.
Real $\Delta K$ leads to $v>0$, which is marginally relevant, making the transition become first-order\cite{Youjin2010, 2016PhRvE..94e2103Q}.
The complex couplings ($\Delta K=i\phi, \Delta K^2<0$) bring the model to the region $v<0$,
which is marginally irrelevant, leading the model to belong to the 2D four-state Potts universality class with logarithmic corrections,  as present in the 2D four-state Potts model\cite{blote2017}.
The corresponding renormalization group (RG) flow diagram is shown in Fig.~\ref{fig:phase_diagram}(b).

We now show that the GBW model is equivalent to a 1D quantum model. Consider a $L_x\times L_y$ triangular lattice wrapped on a torus with circumference $L_x$, see Fig.~\ref{fig:TRI_lattice}. The partition function is then written as
 \begin{equation}
{\cal Z}=
\sum_{\vec{s}_1}\sum_{\vec{s}_2}\ldots\sum_{\vec{s}_{L_y}}T_{\vec{s}_1\vec{s}_2} T_{\vec{s}_2\vec{s}_3}\cdots T_{\vec{s}_{L_y}\vec{s}_1}=
	 \Tr(T^{L_y}),
	 \label{pf_Z}
\end{equation}
where ~$\vec{s}_i$ is a vector representing a row of $L_x$ Ising spins coded as binary digits,
$T_{\vec{s}_i\vec{s}_{i+1}}$ is a $2^{L_x} \times 2^{L_x}$  transfer matrix from the $i$-th row to the $(i+1)$-th row along the $y$ direction:
\begin{equation}
	T_{\vec{s}_i, \vec{s}_{i+1}}=
	\exp{(K_{\rm up}\sum_{\triangle_i} \sigma _{i}\sigma _{j}\sigma_{k}+K_{\rm down} \sum_{\bigtriangledown_i} \sigma _{l}\sigma _{m}\sigma_{n})},
\end{equation}
where $\triangle_i$ ($\bigtriangledown_i$) refers to up-triangle (down triangle) between row $i$ and $i+1$.
It is clear that $T_{j,k}$ becomes $T_{k,j}^*$ after interchanging up- and down triangles, meaning $T$ is hermitian.
With the $y$-direction taken as the quantum ``time" direction, the transfer matrix $T$, which evolves the state from the $i$-th row to the $(i+1)$-th row,
is equivalent to a quantum operator $\exp(-H_Q)$\cite{Sachdev_2011}.
Therefore, the partition function Eq.~(\ref{pf_Z}) can be written as
 \begin{equation}
	 {\cal Z} \sim \Tr \exp{(-H_Q/T_Q)}
\end{equation}
with $T_Q=1/L_y$,
which describes a quantum model with Hamiltonian $H_Q$ at temperature $T_Q \to 0$ at the limit $L_y \to \infty$.
So the GBW model can be viewed as a path integral representation of this quantum 1D model.

In the language of classical representations of the GBW model, the Boltzmann weight of a given
configuration $\Gamma\equiv (\sigma _{1},\sigma _{2},\cdots, \sigma _{N})$ is
\begin{equation}
W(\Gamma, T)=e^ {\left ( K+i\phi  \right )A(\Gamma)+\left ( K-i\phi  \right )B(\Gamma)},
\end{equation}
with $A(\Gamma)=\sum_\triangle \sigma _{i}\sigma _{j}\sigma_{k}$ and $B(\Gamma)=\sum_{\bigtriangledown } \sigma _{l}\sigma _{m}\sigma_{n}$,
$T$ is the temperature.
By binding configuration $\Gamma$ with its $\pi$-rotated counterpart  $\Gamma'$  about a lattice direction (see Fig.~\ref{fig:Binding_configuration}), to form a new
configuration $\tilde{\Gamma}$, the imaginary contributions to the partition function $\cal Z$ of the model (\ref{Eq:Ham}) cancel:% and the partition function becomes
\begin{equation}
	\mathcal{Z}(T)= \sum_{\Gamma} W(\Gamma, T)=\frac{1}{2} \sum_{\tilde{\Gamma}} W(\tilde{\Gamma}, T),
    \label{Zbw}
\end{equation}
with the weight of the bound configuration %$
\begin{equation}
\label{Eq:Wbindingconf}
	\begin{split}
		W(\tilde{\Gamma}, T) &\equiv  W(\Gamma, T)+W(\Gamma', T)\\
		&=2e^{K\left ( A\left ( \Gamma  \right ) +B\left ( \Gamma \right )  \right ) }\cos\left [ \left ( A\left ( \Gamma \right ) -B\left ( \Gamma \right )  \right ) \phi\right ].
	\end{split}
\end{equation}
Clearly, the multiplicative factor $\cos\left [ \left ( A\left ( \Gamma \right ) -B\left ( \Gamma \right )  \right ) \phi\right ] $ in the weight reveals the sign problem in this model.

The standard way to simulate a model with a sign problem is to define a reference model using the absolute values
of the weights $W(\tilde{\Gamma}, T)$
\begin{equation}
    {\cal Z}_+(T)=\frac{1}{2} \sum_{\tilde{\Gamma}} |W(\tilde{\Gamma}, T)|.
    \label{Zbw+}
\end{equation}
The average sign is then defined as
\begin{equation}
\label{Eq:average_sign}
\left \langle S \right \rangle_+ =\frac{\sum_{\tg} \left| W(\tg, T )\right | S(\tg)}
{\sum_{\tg}\left |W(\tg, T)\right|} %=\frac{\sum_{\tg}W(\tg, T)}{\sum_{\tg}\left |W(\tg, T)\right |}
=\frac{{\cal Z}(T)}{{\cal Z}_+(T)},
\end{equation}
where the sign of a configuration $\tg$ reads
\begin{equation}
\label{Eq:sign}
S(\tg)=
\frac{W(\tg, T)}{|W(\tg, T)|}
=\frac{\cos\left [ \left ( A\left ( \Gamma \right ) -B\left ( \Gamma \right )  \right )\phi \right ]}{|\cos\left [ \left ( A\left ( \Gamma \right ) -B\left ( \Gamma \right )  \right ) \phi \right ]|}.
\end{equation}

The four ground states of the proper Baxter-Wu model persist for the GBW model (\ref{Zbw})\cite{Youjin2010, blote2017}.
These ground states can be transformed into each other by flipping Ising spins on two sets of three sublattices,
without changing the product of the three spins in each up- or down-triangle, hence,
the weight Eq.~(\ref{Eq:Wbindingconf}) and its absolute value.
Then the models defined by ${\cal Z}$ and ${\cal Z}_+$ should share the universal behavior as the four-state Potts model, similar to the proper Baxter-Wu model\cite{Domany_BW-Potts}.
Unfortunately, the ${\cal Z}_+$ is not self-dual anymore,  since the sum of absolute weight does not equal the absolute value of the sum of weight.

We will simulate the ${\cal Z}_+$ model (\ref{Zbw+}) using a Metropolis algorithm\cite{Metropolis1953,Binder_2012}.
The linear size of systems is taken as multiples of 3. Periodic boundary conditions are imposed. For each size, $10^8$ samples are taken to calculate the MC averages.
Below, we will study the relationship between average signs and critical behavior.
We will investigate the criticality of the ${\cal Z}_+$ model to show a new approach to studying systems with sign problems.

\section{Results}
\label{sec:results}
\subsection{The average conventional sign near the critical point of the GBW model}
To study the relationship between the average sign and the phase transition in the GBW model,
we choose $K=K_c/T, \phi=\phi_c/T$, where $(K_c=0.3, \phi_c=0.440322696)$ is a self-dual point satisfying Eq.~(\ref{cp}),
thus, the complex couplings $K_{\rm up}=K+i \phi$  and $K_{\rm down}=K-i \phi$  vary with temperature $T$,  with $T_c=1$ the critical point.

\begin{figure}[htb]
  \includegraphics[width=0.5\textwidth]{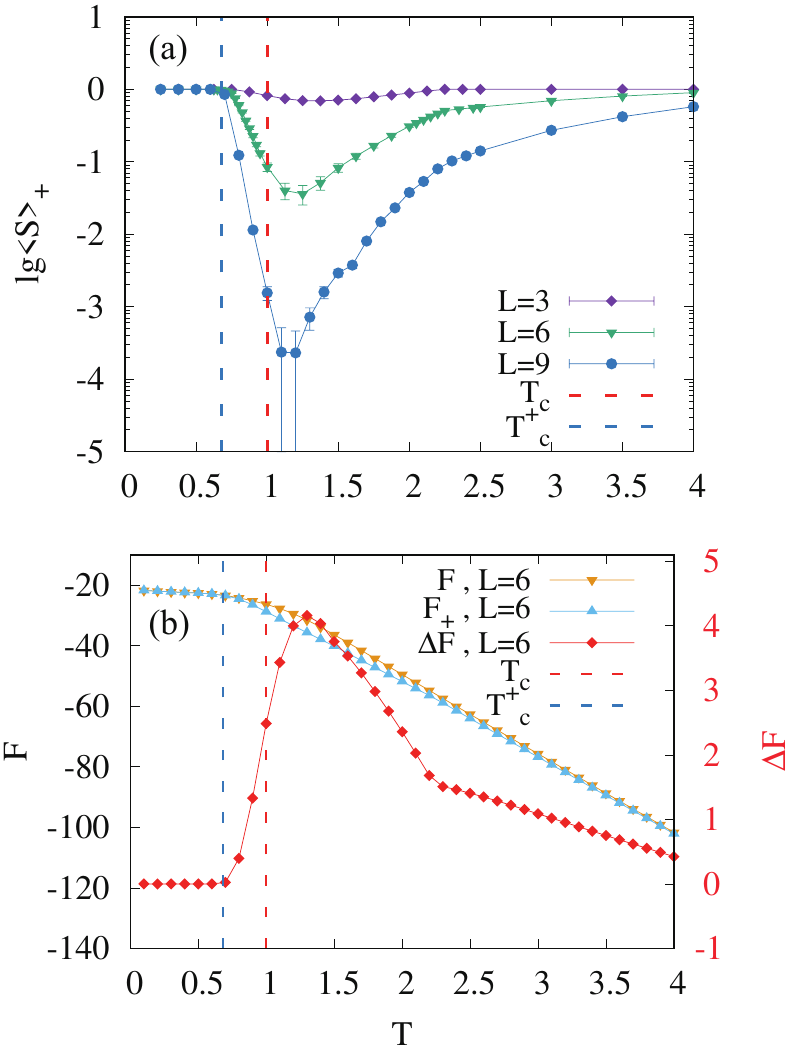}
\caption{\label{fig:sign_T}
The GBW model with $K=K_c/T$, $\phi=\phi_c/T$ where $(K_c, \phi_c)=(0.3, 0.440322696)$. The red dashed line denotes the critical point $T_c=1$ of the original model; The blue dashed line indicates the critical point $T_c^+=0.680738(23)$ of the ${\cal Z}_+$ model.
(a)The average sign $\< S \>_+$ as a function of the temperature $T$ for different system sizes.
(b)The exactly calculated free energies $F$ of the GBW model, $F_+$ of the ${\cal Z}_+$ model, and the difference
$\Delta F$ between them as functions of temperature $T$ for $L=6$.
}
\end{figure}

We calculate the average sign as a function of $T$ using MC simulations to ${\cal Z}_+$. The results for
different system sizes are shown in Fig.~\ref{fig:sign_T}(a).
In the high temperature limit ($T\to \infty$, hence $\phi= \phi_c/T \to  0$), we have $\cos\left [ ( A ( \Gamma ) -B ( \Gamma ) ) \phi \right] \approx 1$ for all configurations; whilst in the low-temperature limit ($T \to 0$), the system is in one of its four ground states, where $A( \Gamma ) -B ( \Gamma )=0$. Consequently, the average sign $\< S \>_+=1$ in both limits.
Between these two limits, $\<S\>_+$ rapidly tends to zero as the system size increases, indicating that  the sign problem in this model is very
severe and makes it difficult to study its critical behavior using the reweighting technique\cite{Troyer2005, 10.21468/SciPostPhysCodeb.1-v2.4}.
Nevertheless, $\ln \< S\>_+$ versus $T$ curves exhibit a minimum closing to the critical point $T_c=1$, in support of the finding in Ref. \cite{Mondaini2022}.

Following Ma et al.\cite{Manvsen2024}, we try to understand this phenomenon by rewriting Eq.~(\ref{Eq:average_sign}) as:
\begin{equation}
\label{Eq:average_sign2}
\langle S \rangle_+ =\exp(-\Delta F/T),
\end{equation}
where $\Delta F=F-F_+$ is the free energy difference of the original and the reference systems,
$F=-T\ln {\cal Z}$ and $F_+=-T \ln {\cal Z}_+$.
As shown in Fig.~\ref{fig:sign_T}(b), $F_+(T)$ and $F(T)$ curves are similar at finite size, but the critical point of the ${\cal Z}+$ model is much smaller than 1 (we shall show that the
critical point of the ${\cal Z}+$ model is  $T_c^+=0.680738(23)$ below) and no singular behavior is expected
near $T=1$, therefore, the peak of the $\Delta F$ versus $T$ curve near the critical point $T_c=1$ may be related
to the phase transition in the original model,
which results in the minimum of $\ln \<S\>_+(T)$ curves shown in Fig.~\ref{fig:sign_T}(a) appearing at roughly the same position.
Unfortunately, it is almost impossible to draw exact conclusions about the
phase transition point based solely on the behavior of the average sign, since determining the minimum of
$\< S \>_+$ accurately requires computational cost that scales exponentially with the system volume.

On the other hand, the average sign $\< S \>_+$ may give a false indication of a phase transition in the current GBW model.

\begin{figure}[htb]
  \includegraphics[width=0.45\textwidth]{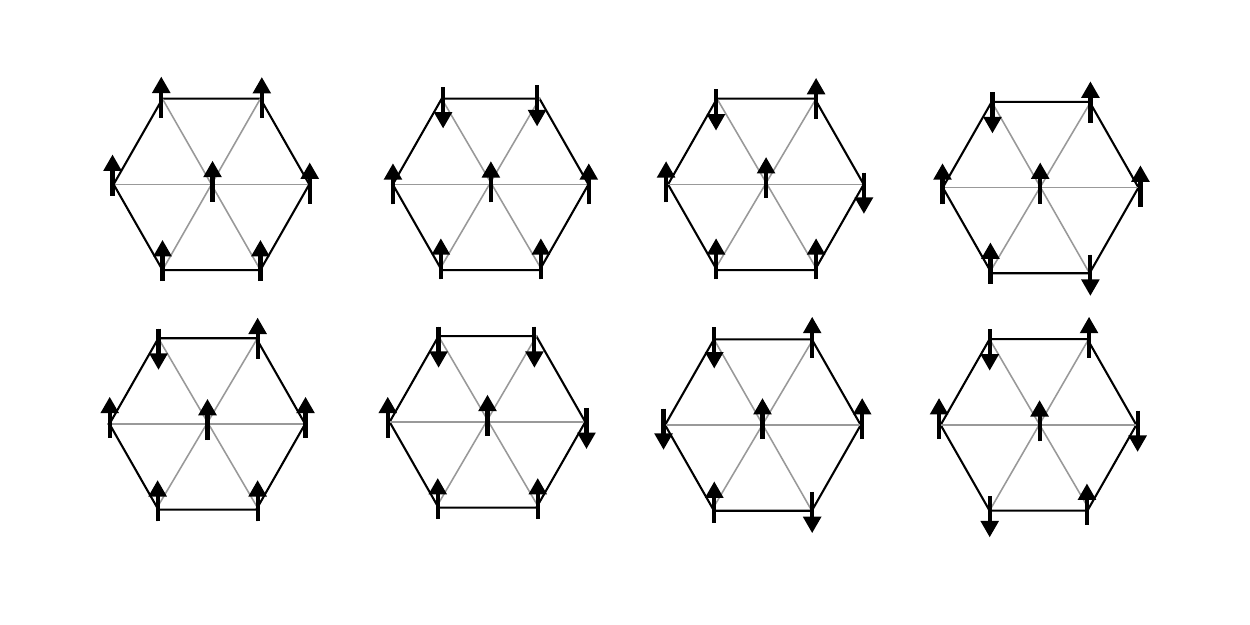}
  \caption{\label{fig:periodicity_lattice}The 8 types of nearest neighbor configurations of a given site on the triangular lattice, each with 2, 12, 12, 6, 6, 12, 12, and 2 symmetric transformations. Flipping the center spin results in
$ |\Delta \left ( A\left ( \Gamma \right ) -B\left ( \Gamma \right )  \right )|= 0, 8, 0, 0, 0, 0, 8, 0$.  This means that the possible value of $|\Delta \left ( A\left ( \Gamma \right ) -B\left ( \Gamma \right )  \right )|$ must be multiples of 8.  }
\end{figure}

\begin{figure}[htb]
    \includegraphics[width=0.45\textwidth]{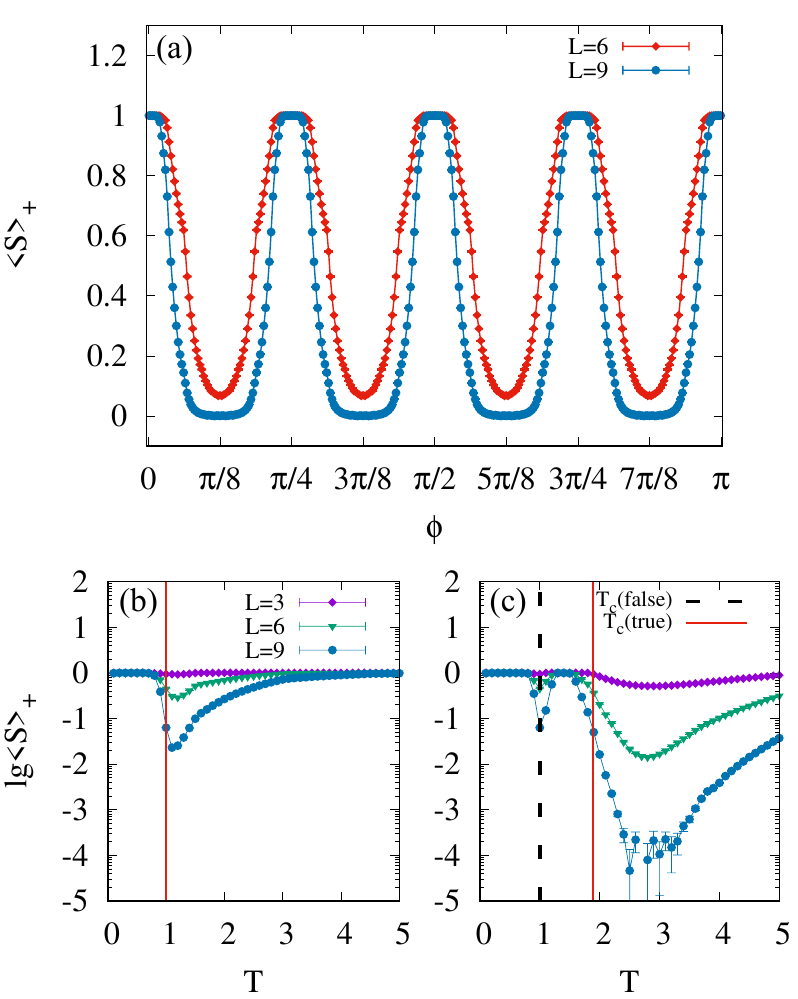}
  \caption{\label{fig:periodicity}
	(a) The average sign $\< S \>_+$ exhibits a periodicity of $\pi/4$ in $\phi$ at fixed $K=0.3$. (b)-(c)  Variation of the average sign $\< S \>_+$ with $T$ along the two paths (see FIG~\ref{fig:phase_diagram}(a)): (b) path $K=K_c/T, \phi=\phi_c/T$, which passes through $(K_c, \phi_c)=(0.376165, 0.3)$. The red solid line denotes the critical temperature $T_c=1$; (c) path $K=K_c/T, \phi=(\phi_c+\pi/4)/T$. The red solid line marks the critical temperature $T_c=1.890276$; % the minimum near $T=1$ is a false signal of a phase transition.
    The local minimum near $T = 1$, indicated by the dashed line, falsely indicates a phase transition. It does not lie on the duality line but is related to the point $(0.376165, 0.3)$ on the self-dual line with a period of $\pi/4$. }
\end{figure}

Combining Eq.~(\ref{Eq:average_sign}) and (\ref{Eq:sign}), we can infer that the average sign is periodic in $\phi$ with period $\frac{\pi}{4}$ for a
given $K$. This is evident by considering a local spin-flip.
Each spin on a triangular lattice has six nearest neighbors, giving $2^6$ possible configurations, which can be grouped into 8 types as shown in Fig.~\ref{fig:periodicity_lattice}.
A local spin-flip can change the value of $|\left ( A\left ( \Gamma \right ) -B\left ( \Gamma \right )  \right )|$ by 0 or 8. Hence, the average sign can be expressed as
\begin{equation}
\label{Eq:average_sign1}
\langle S \rangle_+ =\frac{\sum_{\tg}e^{K\left ( A\left ( \Gamma  \right ) +B\left ( \Gamma \right )  \right ) }\cos\left( 8n(\Gamma)\phi  \right )}{\sum_{\tg}e^{K\left ( A\left ( \Gamma  \right ) +B\left ( \Gamma \right )  \right ) }|\cos\left( 8n(\Gamma)\phi  \right)| }
\end{equation}
with all $n(\Gamma)$ integers, which is periodic in $\phi$ with period $\pi/4$.
This periodicity is verified by calculating the average sign $\< S \>_+ $ as a function of $\phi$ at fixed $K=0.3$ for $L=6$ and $L=9$, as shown in Fig.~\ref{fig:periodicity}~(a), which exhibits the expected periodicity of $\pi/4$ in $\phi$.

Suppose the GBW model is at a critical point $K=K_c, \phi=\phi_c$, with $K_c$ and $\phi_c$ satisfying the self-dual relation Eq.~(\ref{cp}).
The periodicity of $\< S \>_+$ leads to $\< S \>_+(K_c, \phi_c)= \< S \>_+(K_c, \phi_c+\pi/4)$.
However, $K=K_c, \phi=\phi_c+\pi/4$ is not a solution of the self-dual Eq.~(\ref{cp}), and therefore,
$T=1$ is not the critical temperature.
Take the self-dual point $K_c=0.376165, \phi_c=0.3$ as an example.  The critical temperature $T_c=1.890276$ for $K=K_c/T$ and $\phi=(\phi_c+\pi/4)/T$, as shown in Fig.~\ref{fig:phase_diagram}(a).
The $\<S \>_+(T)$ as a function of $T$ for $L=6$ and $L=9$ show minima around $T \sim 1$ and $T\sim 3$, as demonstrated in Fig.~\ref{fig:periodicity}~(b)~and~(c). The minimum associated with $T \sim 1$ is a false indication
of a phase transition.

\subsection{The modified sign at the critical point of the GBW model}
\begin{figure}[htb]
\includegraphics[width=0.52\textwidth]{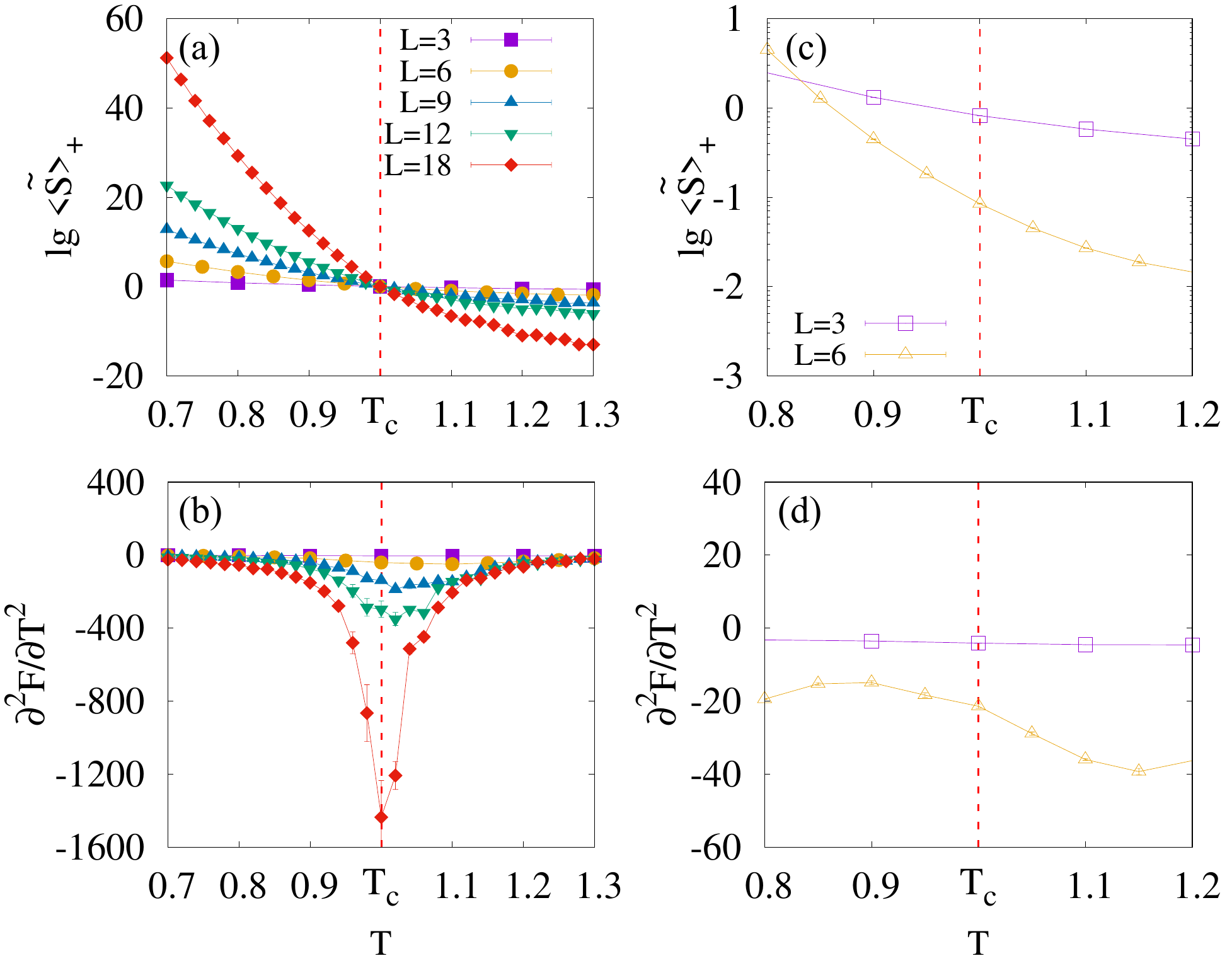}
\caption{\label{fig:modified-sign}
	The average modified sign $\<\widetilde{S}\>^*_+$ and the second-order derivative of the free energy $\partial F^2 / \partial T^2$ as functions of temperature $T$. (a) and (b) for $(K_c, \phi_c)=(0.44,0.031162728)$, (c) and (d)
	for $(0.3,0.440322696)$. }
\end{figure}

The complexity of the average sign in detecting phase transitions can be attributed to the fact
that it is related to both the ${\cal Z}$ model and the ${\cal Z}_+$ model \cite{Manvsen2024}.
To eliminate the influence of the reference ${\cal Z}_+$ system, Ma et al.~\cite{Manvsen2024} proposes a modified sign
\begin{equation}
\label{Eq:sign of BW model}
	\widetilde{S}(\tg, T)= \frac{W(\tilde{\Gamma}, T)}{|W(\tilde{\Gamma},T^*)|},
\end{equation}
and the corresponding average modified sign reads
\begin{equation}
\label{Eq:modified_sign1}
	\<\widetilde{ S }\>^*_+ = \frac{\sum_{\tg} \widetilde{S}(\tg, T) |W(\tg, T^*)|}{\sum_{\tg} |W(\tg, T^*)|}
	=\frac{{\cal Z}(T)}{{\cal Z}_+(T^*)},
\end{equation}

where $T^*$ is a fixed temperature not too far from the critical point of the $\cal Z$ model. In such a way, the modified sign is
directly proportional to ${\cal Z}$.

We now check the effectiveness of the average modified sign in detecting the phase transition of the GBW
model. We choose $T^*=T_c$ and calculate $\<\widetilde{S}\>^*_+$ as function of $T$ with $(K, \phi)$
passing through the self-dual points
$(K_c, \phi_c) =(0.44, 0.031162728)$ and $(0.3, 0.440322696)$, respectively. The results are plotted in
Fig.~\ref{fig:modified-sign}(a) and (c), respectively.

For $(K_c, \phi_c)=(0.44, 0.031162728)$, we see perfect cross of $\<\widetilde{S}\>^*_+(T)$ for different system sizes at $T_c=1$.
This is because the largest possible value of $A(\Gamma)-B(\Gamma)$ in Eq.~(\ref{Eq:Wbindingconf}) for size $L$ is $(A(\Gamma)-B(\Gamma))_{L}
=4[L^2/3-{\rm mod}(L,2)]$. For a given $\phi$, $\cos((A(\Gamma)-B(\Gamma))\phi)>0$ for a range of system sizes $L\le L_{\rm max}$, as far as $\phi$ is
small enough such that $(A(\Gamma)-B(\Gamma))_{L_{\rm max}} \phi \le \pi/2$. For such cases, $\<\widetilde{S}(T, L)\>^*_+=1$ at $T_c$, since ${\cal Z}(T_c)={\cal Z}_+(T_c)$.
In the current case, %$(K_c=0.44, \phi_c=0.031162728),$
we have $L_{\rm max}=6$. However, for $\tg$ with large $A(\Gamma)-B(\Gamma)$, the
value $A(\Gamma)+B(\Gamma)$ is much smaller than other configurations; therefore, it contributes a very small weight in ${\cal Z}_+$.
This results in that $\< \widetilde{S}(T_c, L)\>^*_+=1$ for $L\le 12$. Starting from $L=18$, the
crossings move away from the critical point $T_c=1$.

Nevertheless, from  Eq.~(\ref{Eq:modified_sign1}), we know  $ \ln (\< \widetilde{S}\>_+^*) = -F/T + c$, with $c$ is a constant and $F$ is the free energy
of the original system. It is
then argued \cite{Manvsen2024} that $\< \widetilde{S}\>^*_+(T, L)$ encodes the information of the free energy of the system and therefore describes
the phase transition.
The second-order derivative of the free energy is expressed with derivatives of $\<\widetilde{S}\>^*_+$ to $T$ \cite{Manvsen2024},
%Eq.~(\ref{Eq:freeE3}).
\begin{equation}
	\frac{\partial ^2 F}{\partial T^2} =\frac{\partial^2 (T\ln{\< \widetilde{{S}}\>^*_+)}}{\partial T^2},
\label{Eq:freeE3}
\end{equation}
which is related to the heat capacity, thus should develop a negative peak diverging with system size at the critical point.
The average modified sign is then claimed as a useful probe for the phase transition.
Such a prediction has been tested true by numerically differentiating $\ln \< \widetilde{S}\>^*_+$ of a frustrated $J_1-J_2$ model
with an Ising-like phase transition \cite{Manvsen2024}.

A little more algebra shows that the second-order derivative of the free energy can be written as follows:
\begin{equation}
	\begin{split}
\frac{\partial ^2 F}{\partial T^2} %&=\partial ^2 [T[\< \widetilde{{S}}\>]/\partial T^2
		&=-2\frac{\< {\widetilde{S}' } \>^*_+  } {\< \widetilde{ S }\rangle^*_+ }
		+T\left ( \frac{\< \widetilde{ S }'  \>^*_+ }{\< \widetilde{ S } \>^*_+ } \right)^2
		-T\frac{{\langle \widetilde{S}''  \rangle^*_+ } }{\< \widetilde{ S }  \>^*_+ },
	\end{split}
\label{2nd-derivative}
\end{equation}
where the estimators $\widetilde{S}'(\tg, T)=\partial \widetilde{S}(\tg, T)/\partial T$ and $\widetilde{S}''(\tg,T)=\partial \widetilde{S}'(\tg, T)/\partial T$ are presented in Appendix \ref{appendixa}.
Therefore,  $\partial ^2 F/\partial T^2$  can be calculated by MC simulating the referenced model at $T^*=T_c$.

In such a way, we calculate the second-order derivative of the free energy of the GBW model with $(K_c, \phi_c)=(0.44, 0.031162728)$
near its transition point $T_c=1$.  The results are shown in Fig.~\ref{fig:modified-sign} (b).
We observe diverging negative peaks approaching the critical point $T_c=1$ as system size increases.
This result suggests that the average modified sign is, in principle, a valid probe to detect phase transitions.

However, it is evident that the calculation of the second-order derivative of the free energy using Eq.~(\ref{2nd-derivative}) suffers from exactly
the same difficulty as the reweighting technique.
At $T \sim T_c$, $\< \widetilde{S}\>_+^* \sim \<\widetilde{S}\>_+ \sim \exp(-\Delta F/T_c)$ with $\Delta F$  the free energy difference between the original
model and the reference model. Take the first term
$-2\< {\widetilde{S}' } \>^*_+   /\< \widetilde{ S }\rangle^*_+ $ in Eq.~(\ref{2nd-derivative}) as an example.
Suppose $M$ is the number of sampled independent MC configurations in the simulation of ${\cal Z}_+$, the statistical error
$\epsilon$ of $\< \widetilde{S}'\>^*_+ \sim 1/\sqrt{M}$.  Then, to achieve the same statistical error $\epsilon$
for $-2\< {\widetilde{S}' } \>^*_+  / \< \widetilde{ S }\rangle^*_+ $,   one needs the number of independent MC configurations
$M' \sim M \exp{(2 \Delta F/T_c)}$, which scales exponentially in the volume $L^2$. Similar arguments apply to the other terms in
Eq.~(\ref{2nd-derivative}), resulting in an exponential scaling of MC samples with $L^2$ to achieve a desired statistical error for $\partial ^2 F/\partial T^2$.

In the case $K_c=0.44, \phi_c=0.031162728$,  $\Delta F \sim 0$ up to $L\sim 18$, making  $\partial ^2 F/\partial T^2$ calculatable upto
$L\sim 18$.  However, for $(K_c, \phi_c)=(0.3, 0.440322696)$, $\Delta F \sim 10^3$ for $L =9$ around $T_c=1$, meaning the sign problem becomes much
worse:  $\<S\>_+^*$ and  $\partial ^2 F/\partial T^2$ are reachable only for $L < 9$.  %less than $10^{-3}$.
In principle, for any $(K_c, \phi_c)$ with $\phi_c \neq 0$, $\<S\>^*_+$ and
$\partial ^2 F/\partial T^2$ becomes unreachable quickly with the size $L$ increasing, due to the exponentially increasing requirement of simulation time.

\subsection{The critical behavior of the ${\cal Z}_+$ model}
\begin{figure}[htb] \includegraphics[width=0.5\textwidth]{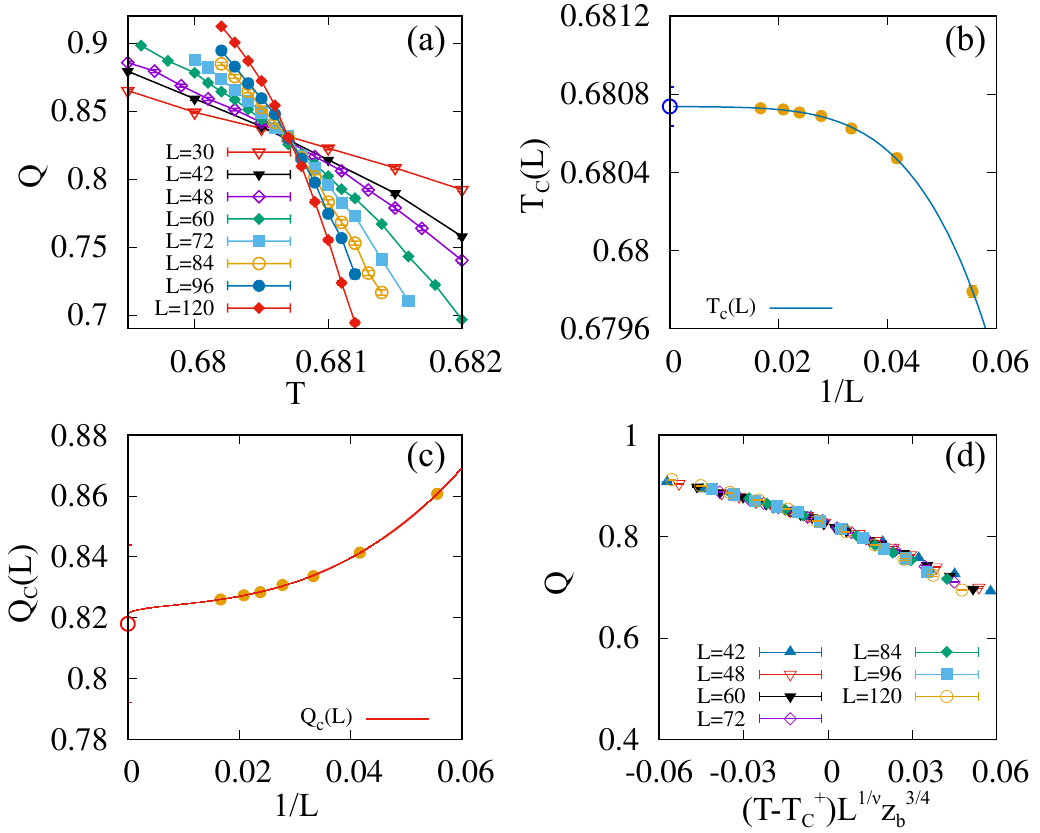}
  \caption{\label{fig:Binder_ratio}
  (a)The Binder ratio $Q$ of the ${\cal Z}_+$ model as function of temperature $T$ with $(K_c, \phi_c)=(0.3, 0.440322696)$.
(b) The phase transition point $T_c^+=0.680738(23)$ is obtained by fitting the crossings
$T_c(L)$ of $Q(L)$ and Q(2L) curves. (c) $Q_c=0.817(26)$ is obtained by fitting
Eq.~(\ref{UcL}) to $Q_c(L)$ with $b=-6$.
    (d) Data collapse with $T^+_c=0.680738,1/\nu=1.5$ and $b=-6$.  }

\end{figure}

The original model defined by ${\cal Z}$ and the reference model defined by ${\cal Z}_+$
should belong to the same universality class as the 2D four-state Potts model, since both
share the same symmetry and ground-state degeneracy as the four-state Potts model \cite{Domany_BW-Potts}.
Therefore, by studying the phase transition in the reference model, we should be able to obtain the critical
behavior of the original GBW model.
To test this anticipation, we now study the critical behavior of the ${\cal Z}_+$ model by performing a finite-size scaling analysis of
observables obtained from our MC simulations.

The order parameter is defined in analogy with the four-state Potts model as
\begin{equation}
\label{Eq:order_parameter}
	m^2=\frac{1}{12}\sum_{i=1}^4\sum_{j=1}^4(\rho_i-\rho_j)^2
\end{equation}
where $i(j)$ denotes four types of triangles: $(+,+,+), (+,-,-), (-,+,-)$, and $(-,-,+),$

associated with the corresponding ground states, and
$\rho_i$ is the density of triangles in the $i$-th ground state.

The fourth-order Binder ratio of $m$ for the ${\cal Z}_+$ model is
\begin{equation}
\label{Eq:Binder_ratio}
	Q(T, L) =\frac{\langle m^2\rangle_+^2}{\langle m^4\rangle_+},
\end{equation}
which is a renormalization invariant that converges to a universal value at the critical point as the size increases.

The marginally irrelevant scaling field $v$ in the 2D four-state Potts model introduces logarithmic corrections to scaling\cite{Cardy_log, Nienhuis_Potts}.
Such a marginally irrelevant scaling field $v\propto -\phi^2$ is present in the GBW model and should introduce similar logarithmic corrections to scaling.
From an RG analysis \cite{Nauenberg_Potts, Cardy_Potts, Salas_Potts}, we can predict the finite-size scaling behavior of the Binder ratio as a function of the temperature scaling field $t\propto T-T_c$, the marginal field $v$, and other irrelevant scaling fields $u$, in the case no magnetic scaling field is present, as
\begin{equation}
	Q(t,v,u,1/L)=Q(tL^{1/\nu}z_v^{3/4},z_v v,uL^{-\omega}, 1)
	\label{scaling}
\end{equation}
where $z_v = 1/(1-(4\pi/\sqrt{3}) v \ln L)$, $\nu=2/3$ is the correlation length exponent of the 2D four-state Potts universality class,
$\omega>0$ is an effective irrelevant exponent.

The scaling function can be expanded in the neighborhood of the critical point
\begin{equation}
	Q(T,L)=Q_c + a_1 (T-T_c^+)L^{1/\nu}z_b^{3/4}+ b_1 z_b + c_1 L^{-\omega}+\dots
\end{equation}
where $z_b=1/(1-b \ln L)$ with $b \propto v$, and $a_1, b_1, c_1$ are unknown constants  \cite{2016PhRvE..94e2103Q}.
Therefore, the crossing temperature $T_c(L)$  of $Q(T, L)$  and $Q(T, 2L)$ curves for size pairs $L$ and $2L$ converge to the critical point $T_c^+$
in the following ways

\begin{equation}
	T_c(L)=T_c^+ + aL^{-1/\nu}z_b^{1/4}+cL^{-\omega-1/\nu}z_b^{-3/4}\ldots
	\label{Tc+}
\end{equation}

The value of the Binder ratio $Q_c(L)$ at $T_c(L)$ converges to $Q_c$ logarithmically,

\begin{equation}
	Q_c(L)=Q_c+az_b+cL^{-\omega}\ldots
	\label{UcL}
\end{equation}

The results of $Q(T, L)$ for several system sizes as a function of temperature are shown in Fig.~\ref{fig:Binder_ratio}(a).
We see that $Q(T)$ for different sizes crosses near $T=0.68$, indicating the presence of a critical point.
The crossing temperature $T_c(L)$  and the crossing Binder ratio $Q_c(L)$ of $Q(T, L)$  and $Q(T, 2L)$ curves for size pairs
$L$ and $2L$ are found by polynomial fitting
the $Q(T,L)$ curves in the neigborhood of $T=0.68$.
The obtained $T_c(L)$ and $Q_c(L)$ vs $1/L$ are illustrated in Fig.~\ref{fig:Binder_ratio}(b) and (c), respectively.

We then fit Eq.~(\ref{Tc+}) to $T_c(L)$ with $1/\nu$ fixed as 3/2 and obtain statistically sound fitting %with reduced $\chi^2=2.4$ and $p-$value $0.06$
for $L\ge L_{\rm min}=18$ data. The fits remain stable upon further excluding small-sized points by gradually increasing $L_{\rm min}$ to 24, indicating Eq.~(\ref{Tc+}) describes the scaling behavior of $T_c(L)$ correctly. In particular, the value of $b$ in $z_b$ is found to be negative, indicating the scaling field $v$ of the model is irrelevant, and leading the model to belong to the 2D four-state Potts universality class. Unfortunately, the fitting result of $b$ carries a large uncertainty, making the specific value meaningless. The best estimate of $T_c$ is 0.6807(1), which is significantly different from the critical point temperature $T_c=1$ of the original model.

The leading logarithmic term in Eq.~(\ref{UcL}) makes the convergence of $Q_c(L)$ extremely slow. Fitting Eq.~(\ref{UcL}) to $Q_c(L)$ yields $Q_C\sim 0.812$ with large
uncertainty.

We have also attempted to fix $b$ to a finite value, e.g., $b=-6$, and find $T_c^+=0.680738(23)$ and $Q_C=0.817(26)$ by fitting Eq.~(\ref{Tc+}) to $T_c(L)$
and fitting Eq.~(\ref{UcL}) to $Q_c(L)$, respectively.
The obtained  $Q_c$ agrees with known values for the 2D four-state Potts universality class \cite{Capponi-BW-Tfield, Weigel-diluteBW}.

Ignoring irrelevant fields, the scaling behavior Eq.~(\ref{scaling}) of $Q$ reduces to the following scaling function
\begin{equation}
\label{Eq：Data—collapse}
	%Q(T,L)=Q((T-T^+_c)L^{1/\nu}(1-b\ln(L))^{-3/4}),
	Q(T,L)=Q((T-T^+_c)L^{1/\nu}z_b^{3/4}).
\end{equation}
Figure~\ref{fig:Binder_ratio}(d) shows that the data $Q(T, L) $ for different system sizes collapse onto the same curve, which is the scaling
function  (\ref{Eq：Data—collapse}), with $T^+_c=0.680738,1/\nu=3/2$ and $b=-6$.
This result both verifies that the phase transition is continuous and suggests  that the
${\cal Z}$ and ${\cal Z}_+$ models belong to the same 2D four-state Potts universality class.

This method, which is based on symmetry analysis and indirectly studying the critical properties of the original model through a reference model ${\cal Z}_+$, offers a novel perspective for investigating systems plagued by sign problems.

\section{Discussions and outlook}
\label{sec:conclusions}

Using the generalized Baxter-Wu model with asymmetric complex coupling--a system with a sign problem yet exactly known
critical properties--as a platform, we have studied the relationship between criticality and both the average sign and the recently proposed
average modified sign.

Our MC simulations have verified that the average sign $\<S\>_+(T)$ curve does have a minimum near
the critical point. However, we have also shown that the curve has an additional minimum at a temperature
away from the critical point, meaning the average sign may give a false indication of a phase transition.

We further studied the effectiveness of the average modified sign in detecting phase
transitions. We confirmed that the modified sign has the influence of the reference system eliminated; therefore, the second-order derivative of the average
modified sign is related to the second-order derivative of the
free energy of the original system. However, we have also demonstrated that detecting a phase transition by
calculating the second-order derivative of the average modified sign is impractical: the presence of the average sign in the denominator of the second derivative of the free energy significantly impedes its effectiveness, due to the exponential scaling of MC cost with system volume.

More importantly, based on symmetry analysis and universality class, we proposed studying the critical properties of the original model through the reference ${\cal Z}_+$ model, defined by the absolute values of the corresponding weights of the original model.
We observed that the reference ${\cal Z}_+$ model of the original GBW model
shares the same groundstate symmetry and degeneracy as the sign-problematic original model. Using finite-size scaling analysis based on MC simulations, we have confirmed that the phase
transition of the ${\cal Z}_+$ model shares the same four-state Potts universality class as the original model.

However, it should be noted that this method is model-dependent.
%For example, the frustrated
%easy plane $J_1-J_2$ model bears such property that the reference ${\cal Z}_+$ model share the same symmetry as the original model, but
For example, for fermionic models, the reference model typically
has different symmetry from the original model, therefore, different universal properties are expected in the reference model.
%Its effects—such as how taking the absolute value of weights alters the physical content and universality class—must be analyzed on a case-by-case basis for each specific sign-problematic system (e.g.,
%In many systems with sign-problems, e.g., fermionic models, frustrated magnets, or gauge-coupled spin models, taking absolute values of the weights

This method provides a novel framework for investigating systems plagued by sign problems.
Applying this approach to study critical properties of various recently proposed \cite{PhysRevLett.133.077101} models with complex couplings is interesting. Work along these lines is currently in progress.

\begin{acknowledgments}
We would like to thank H. Q. Lin and F. F. Assaad for valuable discussions. Y. Ling, Y. Wang, and Y. Liu were supported by the National Natural Science Foundation of China under Grant Nos.~12305039 and 12574251 as well as the Fundamental Research Funds for the Central Universities from the Beijing University of Posts and Telecommunications under Grant  Nos. 2025JCTP08 and 2023RC42 . Y. Ling was supported by the College Student Innovation Training Program Project from Beijing University of Posts and Telecommunications under Grant No.202407010. W. Guo was supported by the National Natural Science Foundation of China under Grant Nos. 12574252 and 12175015. We were also supported by the Open Fund of Key Laboratory of Multiscale Spin Physics 
(Ministry of Education), Beijing Normal University, with No.SPIN2025K03.
\end{acknowledgments}

{\it{\color{blue} Data availability.-}}
The data that support the findings of this article are openly available\cite{liu2026zenodo}.

%\clearpage
\appendix
\section{ Modified sign}
\label{appendixa}
We set $C(\Gamma)= A (\Gamma)+B(\Gamma)$ and $D(\Gamma)= A (\Gamma)-B(\Gamma)$ with
$A(\Gamma)=\sum_\triangle \sigma _{i}\sigma _{j}\sigma_{k}$
%for the sum of up-triangles of $\Gamma$
and $B(\Gamma)=\sum_{\bigtriangledown } \sigma _{l}\sigma _{m}\sigma_{n}$.

%for the sum of down-triangles of $\Gamma$.
The first order derivative is written as:
\begin{eqnarray}
%\begin{equation}
%\begin{split}
	\widetilde {S}'(\tilde{\Gamma}, T) &=& \frac{-e^{\frac{K_c C\left ( \Gamma  \right )}{T}  } } {	T^2  |W(\tg, T^*)|}
 \nonumber\\
 %\\
	&\times &\left [ K_cC (\Gamma) \cos (\frac{\phi _c C(\Gamma)}{T})  -\phi _cD (\Gamma)
\sin(\frac{\phi_cD(\Gamma)}{T})  \right ] .
	%{|2e^{K\left (C\left ( \Gamma  \right ) )  \right ) }\cos\left [ \left ( D\left ( \Gamma \right ))  \right ) \phi\right ]|}
   %\nonumber\\
%   \end{split}
%  \end{equation}
\end{eqnarray}

The second-order derivative is written as
\begin{equation}
    \begin{split}
	    & \widetilde{S}''(\tilde{\Gamma}, T) =
        \frac{e^{\frac{K_c}{T}C(\Gamma)}}{T^4 |W(\tg, T^*)|}
        \Bigg\{ \left[2T + K_c D(\Gamma)\right]
        \\
        &\times \left[K_c C(\Gamma) \cos(\frac{\phi_c}{T} D(\Gamma))
        - \phi_c D(\Gamma) \sin (\frac{\phi_c}{T} D(\Gamma))\right]
         \\
        &- \phi_c D(\Gamma)
        \\
	&\times    \left[ K_c C(\Gamma) \sin(\frac{\phi_c}{T} D(\Gamma))
	     + \phi_c D(\Gamma) \cos(\frac{\phi_c}{T} D(\Gamma)) \right]
	    \Bigg\}.
    \end{split}
	\label{d2SdT2}
\end{equation}
\clearpage

\bibliography{fassaad}

\begin{thebibliography}{31}%
\makeatletter
\providecommand \@ifxundefined [1]{%
 \@ifx{#1\undefined}
}%
\providecommand \@ifnum [1]{%
 \ifnum #1\expandafter \@firstoftwo
 \else \expandafter \@secondoftwo
 \fi
}%
\providecommand \@ifx [1]{%
 \ifx #1\expandafter \@firstoftwo
 \else \expandafter \@secondoftwo
 \fi
}%
\providecommand \natexlab [1]{#1}%
\providecommand \enquote  [1]{``#1''}%
\providecommand \bibnamefont  [1]{#1}%
\providecommand \bibfnamefont [1]{#1}%
\providecommand \citenamefont [1]{#1}%
\providecommand \href@noop [0]{\@secondoftwo}%
\providecommand \href [0]{\begingroup \@sanitize@url \@href}%
\providecommand \@href[1]{\@@startlink{#1}\@@href}%
\providecommand \@@href[1]{\endgroup#1\@@endlink}%
\providecommand \@sanitize@url [0]{\catcode `\\12\catcode `\$12\catcode
  `\&12\catcode `\#12\catcode `\^12\catcode `\_12\catcode `\%12\relax}%
\providecommand \@@startlink[1]{}%
\providecommand \@@endlink[0]{}%
\providecommand \url  [0]{\begingroup\@sanitize@url \@url }%
\providecommand \@url [1]{\endgroup\@href {#1}{\urlprefix }}%
\providecommand \urlprefix  [0]{URL }%
\providecommand \Eprint [0]{\href }%
\providecommand \doibase [0]{https://doi.org/}%
\providecommand \selectlanguage [0]{\@gobble}%
\providecommand \bibinfo  [0]{\@secondoftwo}%
\providecommand \bibfield  [0]{\@secondoftwo}%
\providecommand \translation [1]{[#1]}%
\providecommand \BibitemOpen [0]{}%
\providecommand \bibitemStop [0]{}%
\providecommand \bibitemNoStop [0]{.\EOS\space}%
\providecommand \EOS [0]{\spacefactor3000\relax}%
\providecommand \BibitemShut  [1]{\csname bibitem#1\endcsname}%
\let\auto@bib@innerbib\@empty
%</preamble>
\bibitem [{\citenamefont {Troyer}\ and\ \citenamefont
  {Wiese}(2005)}]{Troyer2005}%
  \BibitemOpen
  \bibfield  {author} {\bibinfo {author} {\bibfnamefont {M.}~\bibnamefont
  {Troyer}}\ and\ \bibinfo {author} {\bibfnamefont {U.-J.}\ \bibnamefont
  {Wiese}},\ }\bibfield  {title} {\bibinfo {title} {Computational complexity
  and fundamental limitations to fermionic quantum monte carlo simulations},\
  }\href {https://doi.org/10.1103/PhysRevLett.94.170201} {\bibfield  {journal}
  {\bibinfo  {journal} {Phys. Rev. Lett.}\ }\textbf {\bibinfo {volume} {94}},\
  \bibinfo {pages} {170201} (\bibinfo {year} {2005})}\BibitemShut {NoStop}%
\bibitem [{\citenamefont {Assaad}\ \emph {et~al.}(2025)\citenamefont {Assaad},
  \citenamefont {Bercx}, \citenamefont {Goth}, \citenamefont {Götz},
  \citenamefont {Hofmann}, \citenamefont {Huffman}, \citenamefont {Liu},
  \citenamefont {Toldin}, \citenamefont {Portela},\ and\ \citenamefont
  {Schwab}}]{10.21468/SciPostPhysCodeb.1-v2.4}%
  \BibitemOpen
  \bibfield  {author} {\bibinfo {author} {\bibfnamefont {F.~F.}\ \bibnamefont
  {Assaad}}, \bibinfo {author} {\bibfnamefont {M.}~\bibnamefont {Bercx}},
  \bibinfo {author} {\bibfnamefont {F.}~\bibnamefont {Goth}}, \bibinfo {author}
  {\bibfnamefont {A.}~\bibnamefont {Götz}}, \bibinfo {author} {\bibfnamefont
  {J.~S.}\ \bibnamefont {Hofmann}}, \bibinfo {author} {\bibfnamefont
  {E.}~\bibnamefont {Huffman}}, \bibinfo {author} {\bibfnamefont
  {Z.}~\bibnamefont {Liu}}, \bibinfo {author} {\bibfnamefont {F.~P.}\
  \bibnamefont {Toldin}}, \bibinfo {author} {\bibfnamefont {J.~S.~E.}\
  \bibnamefont {Portela}},\ and\ \bibinfo {author} {\bibfnamefont
  {J.}~\bibnamefont {Schwab}},\ }\bibfield  {title} {\bibinfo {title} {{The ALF
  (Algorithms for Lattice Fermions) project release 2.4. Documentation for the
  auxiliary-field quantum Monte Carlo code}},\ }\href
  {https://doi.org/10.21468/SciPostPhysCodeb.1-v2.4} {\bibfield  {journal}
  {\bibinfo  {journal} {SciPost Phys. Codebases}\ ,\ \bibinfo {pages} {1}}
  (\bibinfo {year} {2025})}\BibitemShut {NoStop}%
\bibitem [{\citenamefont {Li}\ \emph {et~al.}(2015)\citenamefont {Li},
  \citenamefont {Jiang},\ and\ \citenamefont {Yao}}]{Zi-Xiang2015}%
  \BibitemOpen
  \bibfield  {author} {\bibinfo {author} {\bibfnamefont {Z.-X.}\ \bibnamefont
  {Li}}, \bibinfo {author} {\bibfnamefont {Y.-F.}\ \bibnamefont {Jiang}},\ and\
  \bibinfo {author} {\bibfnamefont {H.}~\bibnamefont {Yao}},\ }\bibfield
  {title} {\bibinfo {title} {Solving the fermion sign problem in quantum monte
  carlo simulations by majorana representation},\ }\href
  {https://doi.org/10.1103/PhysRevB.91.241117} {\bibfield  {journal} {\bibinfo
  {journal} {Phys. Rev. B}\ }\textbf {\bibinfo {volume} {91}},\ \bibinfo
  {pages} {241117} (\bibinfo {year} {2015})}\BibitemShut {NoStop}%
\bibitem [{\citenamefont {Chang}\ and\ \citenamefont
  {Li}(2023)}]{chang2023boosting}%
  \BibitemOpen
  \bibfield  {author} {\bibinfo {author} {\bibfnamefont {W.-X.}\ \bibnamefont
  {Chang}}\ and\ \bibinfo {author} {\bibfnamefont {Z.-X.}\ \bibnamefont {Li}},\
  }\href@noop {} {\bibinfo {title} {Boosting quantum monte carlo and
  alleviating sign problem by gutzwiller projection}} (\bibinfo {year}
  {2023}),\ \Eprint {https://arxiv.org/abs/2303.13403} {arXiv:2303.13403
  [cond-mat.str-el]} \BibitemShut {NoStop}%
\bibitem [{\citenamefont {Karakuzu}\ \emph {et~al.}(2023)\citenamefont
  {Karakuzu}, \citenamefont {Cohen-Stead}, \citenamefont {Batista},
  \citenamefont {Johnston},\ and\ \citenamefont {Barros}}]{Karakuzu2023}%
  \BibitemOpen
  \bibfield  {author} {\bibinfo {author} {\bibfnamefont {S.}~\bibnamefont
  {Karakuzu}}, \bibinfo {author} {\bibfnamefont {B.}~\bibnamefont
  {Cohen-Stead}}, \bibinfo {author} {\bibfnamefont {C.~D.}\ \bibnamefont
  {Batista}}, \bibinfo {author} {\bibfnamefont {S.}~\bibnamefont {Johnston}},\
  and\ \bibinfo {author} {\bibfnamefont {K.}~\bibnamefont {Barros}},\
  }\bibfield  {title} {\bibinfo {title} {Flexible class of exact
  hubbard-stratonovich transformations},\ }\href
  {https://doi.org/10.1103/PhysRevE.107.055301} {\bibfield  {journal} {\bibinfo
   {journal} {Phys. Rev. E}\ }\textbf {\bibinfo {volume} {107}},\ \bibinfo
  {pages} {055301} (\bibinfo {year} {2023})}\BibitemShut {NoStop}%
\bibitem [{\citenamefont {Wessel}\ \emph {et~al.}(2017)\citenamefont {Wessel},
  \citenamefont {Normand}, \citenamefont {Mila},\ and\ \citenamefont
  {Honecker}}]{Stefan2017}%
  \BibitemOpen
  \bibfield  {author} {\bibinfo {author} {\bibfnamefont {S.}~\bibnamefont
  {Wessel}}, \bibinfo {author} {\bibfnamefont {B.}~\bibnamefont {Normand}},
  \bibinfo {author} {\bibfnamefont {F.}~\bibnamefont {Mila}},\ and\ \bibinfo
  {author} {\bibfnamefont {A.}~\bibnamefont {Honecker}},\ }\bibfield  {title}
  {\bibinfo {title} {{Efficient Quantum Monte Carlo simulations of highly
  frustrated magnets: the frustrated spin-1/2 ladder}},\ }\href
  {https://doi.org/10.21468/SciPostPhys.3.1.005} {\bibfield  {journal}
  {\bibinfo  {journal} {SciPost Phys.}\ }\textbf {\bibinfo {volume} {3}},\
  \bibinfo {pages} {005} (\bibinfo {year} {2017})}\BibitemShut {NoStop}%
\bibitem [{\citenamefont {Wei}\ \emph {et~al.}(2016)\citenamefont {Wei},
  \citenamefont {Wu}, \citenamefont {Li}, \citenamefont {Zhang},\ and\
  \citenamefont {Xiang}}]{XiangT2016}%
  \BibitemOpen
  \bibfield  {author} {\bibinfo {author} {\bibfnamefont {Z.~C.}\ \bibnamefont
  {Wei}}, \bibinfo {author} {\bibfnamefont {C.}~\bibnamefont {Wu}}, \bibinfo
  {author} {\bibfnamefont {Y.}~\bibnamefont {Li}}, \bibinfo {author}
  {\bibfnamefont {S.}~\bibnamefont {Zhang}},\ and\ \bibinfo {author}
  {\bibfnamefont {T.}~\bibnamefont {Xiang}},\ }\bibfield  {title} {\bibinfo
  {title} {Majorana positivity and the fermion sign problem of quantum monte
  carlo simulations},\ }\href {https://doi.org/10.1103/PhysRevLett.116.250601}
  {\bibfield  {journal} {\bibinfo  {journal} {Phys. Rev. Lett.}\ }\textbf
  {\bibinfo {volume} {116}},\ \bibinfo {pages} {250601} (\bibinfo {year}
  {2016})}\BibitemShut {NoStop}%
\bibitem [{\citenamefont {{Mondaini}}\ \emph {et~al.}(2022)\citenamefont
  {{Mondaini}}, \citenamefont {{Tarat}},\ and\ \citenamefont
  {{Scalettar}}}]{Mondaini2022}%
  \BibitemOpen
  \bibfield  {author} {\bibinfo {author} {\bibfnamefont {R.}~\bibnamefont
  {{Mondaini}}}, \bibinfo {author} {\bibfnamefont {S.}~\bibnamefont
  {{Tarat}}},\ and\ \bibinfo {author} {\bibfnamefont {R.~T.}\ \bibnamefont
  {{Scalettar}}},\ }\bibfield  {title} {\bibinfo {title} {{Quantum critical
  points and the sign problem}},\ }\href
  {https://doi.org/10.1126/science.abg9299} {\bibfield  {journal} {\bibinfo
  {journal} {Science}\ }\textbf {\bibinfo {volume} {375}},\ \bibinfo {pages}
  {418} (\bibinfo {year} {2022})},\ \Eprint {https://arxiv.org/abs/2108.08974}
  {arXiv:2108.08974 [cond-mat.str-el]} \BibitemShut {NoStop}%
\bibitem [{\citenamefont {Mondaini}\ \emph {et~al.}(2023)\citenamefont
  {Mondaini}, \citenamefont {Tarat},\ and\ \citenamefont
  {Scalettar}}]{Mondaini2023}%
  \BibitemOpen
  \bibfield  {author} {\bibinfo {author} {\bibfnamefont {R.}~\bibnamefont
  {Mondaini}}, \bibinfo {author} {\bibfnamefont {S.}~\bibnamefont {Tarat}},\
  and\ \bibinfo {author} {\bibfnamefont {R.~T.}\ \bibnamefont {Scalettar}},\
  }\bibfield  {title} {\bibinfo {title} {Universality and critical exponents of
  the fermion sign problem},\ }\href
  {https://doi.org/10.1103/PhysRevB.107.245144} {\bibfield  {journal} {\bibinfo
   {journal} {Phys. Rev. B}\ }\textbf {\bibinfo {volume} {107}},\ \bibinfo
  {pages} {245144} (\bibinfo {year} {2023})}\BibitemShut {NoStop}%
\bibitem [{\citenamefont {Pan}\ \emph {et~al.}(2023)\citenamefont {Pan},
  \citenamefont {Ma},\ and\ \citenamefont {Ma}}]{TianxingMa2023}%
  \BibitemOpen
  \bibfield  {author} {\bibinfo {author} {\bibfnamefont {Y.}~\bibnamefont
  {Pan}}, \bibinfo {author} {\bibfnamefont {R.}~\bibnamefont {Ma}},\ and\
  \bibinfo {author} {\bibfnamefont {T.}~\bibnamefont {Ma}},\ }\bibfield
  {title} {\bibinfo {title} {Strong ferromagnetic fluctuations in a doped
  checkerboard lattice},\ }\href {https://doi.org/10.1103/PhysRevB.107.245126}
  {\bibfield  {journal} {\bibinfo  {journal} {Phys. Rev. B}\ }\textbf {\bibinfo
  {volume} {107}},\ \bibinfo {pages} {245126} (\bibinfo {year}
  {2023})}\BibitemShut {NoStop}%
\bibitem [{\citenamefont {Ma}\ \emph {et~al.}(2024)\citenamefont {Ma},
  \citenamefont {Sun}, \citenamefont {Pan}, \citenamefont {Cheng},\ and\
  \citenamefont {Yan}}]{Manvsen2024}%
  \BibitemOpen
  \bibfield  {author} {\bibinfo {author} {\bibfnamefont {N.}~\bibnamefont
  {Ma}}, \bibinfo {author} {\bibfnamefont {J.-S.}\ \bibnamefont {Sun}},
  \bibinfo {author} {\bibfnamefont {G.}~\bibnamefont {Pan}}, \bibinfo {author}
  {\bibfnamefont {C.}~\bibnamefont {Cheng}},\ and\ \bibinfo {author}
  {\bibfnamefont {Z.}~\bibnamefont {Yan}},\ }\bibfield  {title} {\bibinfo
  {title} {Defining a universal sign to strictly probe a phase transition},\
  }\href {https://doi.org/10.1103/PhysRevB.110.125141} {\bibfield  {journal}
  {\bibinfo  {journal} {Phys. Rev. B}\ }\textbf {\bibinfo {volume} {110}},\
  \bibinfo {pages} {125141} (\bibinfo {year} {2024})}\BibitemShut {NoStop}%
\bibitem [{\citenamefont {Wood}\ and\ \citenamefont
  {Griffiths}(1972)}]{D_W_Wood_1972}%
  \BibitemOpen
  \bibfield  {author} {\bibinfo {author} {\bibfnamefont {D.~W.}\ \bibnamefont
  {Wood}}\ and\ \bibinfo {author} {\bibfnamefont {H.~P.}\ \bibnamefont
  {Griffiths}},\ }\bibfield  {title} {\bibinfo {title} {A self dual relation
  for an ising model with triplet interactions},\ }\href
  {https://doi.org/10.1088/0022-3719/5/18/001} {\bibfield  {journal} {\bibinfo
  {journal} {Journal of Physics C: Solid State Physics}\ }\textbf {\bibinfo
  {volume} {5}},\ \bibinfo {pages} {L253} (\bibinfo {year} {1972})}\BibitemShut
  {NoStop}%
\bibitem [{\citenamefont {Baxter}\ and\ \citenamefont {Wu}(1973)}]{Baxter1973}%
  \BibitemOpen
  \bibfield  {author} {\bibinfo {author} {\bibfnamefont {R.~J.}\ \bibnamefont
  {Baxter}}\ and\ \bibinfo {author} {\bibfnamefont {F.~Y.}\ \bibnamefont
  {Wu}},\ }\bibfield  {title} {\bibinfo {title} {Exact solution of an ising
  model with three-spin interactions on a triangular lattice},\ }\href
  {https://doi.org/10.1103/PhysRevLett.31.1294} {\bibfield  {journal} {\bibinfo
   {journal} {Phys. Rev. Lett.}\ }\textbf {\bibinfo {volume} {31}},\ \bibinfo
  {pages} {1294} (\bibinfo {year} {1973})}\BibitemShut {NoStop}%
\bibitem [{\citenamefont {Baxter}(1973)}]{Baxter_1973}%
  \BibitemOpen
  \bibfield  {author} {\bibinfo {author} {\bibfnamefont {R.~J.}\ \bibnamefont
  {Baxter}},\ }\bibfield  {title} {\bibinfo {title} {Potts model at the
  critical temperature},\ }\href {https://doi.org/10.1088/0022-3719/6/23/005}
  {\bibfield  {journal} {\bibinfo  {journal} {Journal of Physics C: Solid State
  Physics}\ }\textbf {\bibinfo {volume} {6}},\ \bibinfo {pages} {L445}
  (\bibinfo {year} {1973})}\BibitemShut {NoStop}%
\bibitem [{\citenamefont {Wu}(1982)}]{Wu1982}%
  \BibitemOpen
  \bibfield  {author} {\bibinfo {author} {\bibfnamefont {F.~Y.}\ \bibnamefont
  {Wu}},\ }\bibfield  {title} {\bibinfo {title} {The potts model},\ }\href
  {https://doi.org/10.1103/RevModPhys.54.235} {\bibfield  {journal} {\bibinfo
  {journal} {Rev. Mod. Phys.}\ }\textbf {\bibinfo {volume} {54}},\ \bibinfo
  {pages} {235} (\bibinfo {year} {1982})}\BibitemShut {NoStop}%
\bibitem [{\citenamefont {Domany}\ and\ \citenamefont
  {Riedel}(1978)}]{Domany_BW-Potts}%
  \BibitemOpen
  \bibfield  {author} {\bibinfo {author} {\bibfnamefont {E.}~\bibnamefont
  {Domany}}\ and\ \bibinfo {author} {\bibfnamefont {E.~K.}\ \bibnamefont
  {Riedel}},\ }\bibfield  {title} {\bibinfo {title} {Phase transitions in
  two‐dimensional systems},\ }\href {https://doi.org/10.1063/1.325029}
  {\bibfield  {journal} {\bibinfo  {journal} {Journal of Applied Physics}\
  }\textbf {\bibinfo {volume} {49}},\ \bibinfo {pages} {1315} (\bibinfo {year}
  {1978})},\ \Eprint
  {https://arxiv.org/abs/https://pubs.aip.org/aip/jap/article-pdf/49/3/1315/18379074/1315\_1\_online.pdf}
  {https://pubs.aip.org/aip/jap/article-pdf/49/3/1315/18379074/1315\_1\_online.pdf}
  \BibitemShut {NoStop}%
\bibitem [{\citenamefont {{Deng}}\ \emph {et~al.}(2010)\citenamefont {{Deng}},
  \citenamefont {{Guo}}, \citenamefont {{Heringa}}, \citenamefont
  {{Bl{\"o}te}},\ and\ \citenamefont {{Nienhuis}}}]{Youjin2010}%
  \BibitemOpen
  \bibfield  {author} {\bibinfo {author} {\bibfnamefont {Y.}~\bibnamefont
  {{Deng}}}, \bibinfo {author} {\bibfnamefont {W.}~\bibnamefont {{Guo}}},
  \bibinfo {author} {\bibfnamefont {J.~R.}\ \bibnamefont {{Heringa}}}, \bibinfo
  {author} {\bibfnamefont {H.~W.~J.}\ \bibnamefont {{Bl{\"o}te}}},\ and\
  \bibinfo {author} {\bibfnamefont {B.}~\bibnamefont {{Nienhuis}}},\ }\bibfield
   {title} {\bibinfo {title} {{Phase transitions in self-dual generalizations
  of the Baxter-Wu model}},\ }\href
  {https://doi.org/10.1016/j.nuclphysb.2009.10.014} {\bibfield  {journal}
  {\bibinfo  {journal} {Nuclear Physics B}\ }\textbf {\bibinfo {volume}
  {827}},\ \bibinfo {pages} {406} (\bibinfo {year} {2010})},\ \Eprint
  {https://arxiv.org/abs/0909.0994} {arXiv:0909.0994 [cond-mat.stat-mech]}
  \BibitemShut {NoStop}%
\bibitem [{\citenamefont {Blöte}\ \emph {et~al.}(2017)\citenamefont {Blöte},
  \citenamefont {Guo},\ and\ \citenamefont {Nightingale}}]{blote2017}%
  \BibitemOpen
  \bibfield  {author} {\bibinfo {author} {\bibfnamefont {H.~W.~J.}\
  \bibnamefont {Blöte}}, \bibinfo {author} {\bibfnamefont {W.}~\bibnamefont
  {Guo}},\ and\ \bibinfo {author} {\bibfnamefont {M.~P.}\ \bibnamefont
  {Nightingale}},\ }\bibfield  {title} {\bibinfo {title} {Scaling in the
  vicinity of the four-state potts fixed point},\ }\href
  {https://doi.org/10.1088/1751-8121/aa7b53} {\bibfield  {journal} {\bibinfo
  {journal} {Journal of Physics A: Mathematical and Theoretical}\ }\textbf
  {\bibinfo {volume} {50}},\ \bibinfo {pages} {324001} (\bibinfo {year}
  {2017})}\BibitemShut {NoStop}%
\bibitem [{\citenamefont {{Qian}}\ \emph {et~al.}(2016)\citenamefont {{Qian}},
  \citenamefont {{Deng}}, \citenamefont {{Liu}}, \citenamefont {{Guo}},\ and\
  \citenamefont {{Bl{\"o}te}}}]{2016PhRvE..94e2103Q}%
  \BibitemOpen
  \bibfield  {author} {\bibinfo {author} {\bibfnamefont {X.}~\bibnamefont
  {{Qian}}}, \bibinfo {author} {\bibfnamefont {Y.}~\bibnamefont {{Deng}}},
  \bibinfo {author} {\bibfnamefont {Y.}~\bibnamefont {{Liu}}}, \bibinfo
  {author} {\bibfnamefont {W.}~\bibnamefont {{Guo}}},\ and\ \bibinfo {author}
  {\bibfnamefont {H.~W.~J.}\ \bibnamefont {{Bl{\"o}te}}},\ }\bibfield  {title}
  {\bibinfo {title} {{Equivalent-neighbor Potts models in two dimensions}},\
  }\href {https://doi.org/10.1103/PhysRevE.94.052103} {\bibfield  {journal}
  {\bibinfo  {journal} {\pre}\ }\textbf {\bibinfo {volume} {94}},\ \bibinfo
  {eid} {052103} (\bibinfo {year} {2016})},\ \Eprint
  {https://arxiv.org/abs/1609.08831} {arXiv:1609.08831 [cond-mat.stat-mech]}
  \BibitemShut {NoStop}%
\bibitem [{\citenamefont {Sachdev}(2011)}]{Sachdev_2011}%
  \BibitemOpen
  \bibfield  {author} {\bibinfo {author} {\bibfnamefont {S.}~\bibnamefont
  {Sachdev}},\ }\href@noop {} {\emph {\bibinfo {title} {Quantum Phase
  Transitions}}}\ (\bibinfo  {publisher} {Cambridge University Press,
  Cambridge, UK},\ \bibinfo {year} {2011})\BibitemShut {NoStop}%
\bibitem [{\citenamefont {Metropolis}\ \emph {et~al.}(1953)\citenamefont
  {Metropolis}, \citenamefont {Rosenbluth}, \citenamefont {Rosenbluth},
  \citenamefont {Teller},\ and\ \citenamefont {Teller}}]{Metropolis1953}%
  \BibitemOpen
  \bibfield  {author} {\bibinfo {author} {\bibfnamefont {N.}~\bibnamefont
  {Metropolis}}, \bibinfo {author} {\bibfnamefont {A.~W.}\ \bibnamefont
  {Rosenbluth}}, \bibinfo {author} {\bibfnamefont {M.~N.}\ \bibnamefont
  {Rosenbluth}}, \bibinfo {author} {\bibfnamefont {A.~H.}\ \bibnamefont
  {Teller}},\ and\ \bibinfo {author} {\bibfnamefont {E.}~\bibnamefont
  {Teller}},\ }\bibfield  {title} {\bibinfo {title} {Equation of state
  calculations by fast computing machines},\ }\href
  {https://doi.org/10.1063/1.1699114} {\bibfield  {journal} {\bibinfo
  {journal} {The Journal of Chemical Physics}\ }\textbf {\bibinfo {volume}
  {21}},\ \bibinfo {pages} {1087} (\bibinfo {year} {1953})}\BibitemShut
  {NoStop}%
\bibitem [{\citenamefont {Landau}\ and\ \citenamefont
  {Binder}(2021)}]{Binder_2012}%
  \BibitemOpen
  \bibfield  {author} {\bibinfo {author} {\bibfnamefont {D.~P.}\ \bibnamefont
  {Landau}}\ and\ \bibinfo {author} {\bibfnamefont {K.}~\bibnamefont
  {Binder}},\ }\href
  {https://www.cambridge.org/core/books/guide-to-monte-carlo-simulations-in-statistical-physics/8AE32744A1EBA8EEBC4E5BC50732B15A}
  {\emph {\bibinfo {title} {A Guide to Monte Carlo Simulations in Statistical
  Physics}}},\ \bibinfo {edition} {5th}\ ed.\ (\bibinfo  {publisher} {Cambridge
  University Press},\ \bibinfo {year} {2021})\BibitemShut {NoStop}%
\bibitem [{\citenamefont {Cardy}(1986)}]{Cardy_log}%
  \BibitemOpen
  \bibfield  {author} {\bibinfo {author} {\bibfnamefont {J.~L.}\ \bibnamefont
  {Cardy}},\ }\bibfield  {title} {\bibinfo {title} {Logarithmic corrections to
  finite-size scaling in strips},\ }\href
  {https://doi.org/10.1088/0305-4470/19/17/008} {\bibfield  {journal} {\bibinfo
   {journal} {Journal of Physics A: Mathematical and General}\ }\textbf
  {\bibinfo {volume} {19}},\ \bibinfo {pages} {L1093} (\bibinfo {year}
  {1986})}\BibitemShut {NoStop}%
\bibitem [{\citenamefont {Nienhuis}\ \emph {et~al.}(1979)\citenamefont
  {Nienhuis}, \citenamefont {Berker}, \citenamefont {Riedel},\ and\
  \citenamefont {Schick}}]{Nienhuis_Potts}%
  \BibitemOpen
  \bibfield  {author} {\bibinfo {author} {\bibfnamefont {B.}~\bibnamefont
  {Nienhuis}}, \bibinfo {author} {\bibfnamefont {A.~N.}\ \bibnamefont
  {Berker}}, \bibinfo {author} {\bibfnamefont {E.~K.}\ \bibnamefont {Riedel}},\
  and\ \bibinfo {author} {\bibfnamefont {M.}~\bibnamefont {Schick}},\
  }\bibfield  {title} {\bibinfo {title} {First- and second-order phase
  transitions in potts models: Renormalization-group solution},\ }\href
  {https://doi.org/10.1103/PhysRevLett.43.737} {\bibfield  {journal} {\bibinfo
  {journal} {Phys. Rev. Lett.}\ }\textbf {\bibinfo {volume} {43}},\ \bibinfo
  {pages} {737} (\bibinfo {year} {1979})}\BibitemShut {NoStop}%
\bibitem [{\citenamefont {Nauenberg}\ and\ \citenamefont
  {Scalapino}(1980)}]{Nauenberg_Potts}%
  \BibitemOpen
  \bibfield  {author} {\bibinfo {author} {\bibfnamefont {M.}~\bibnamefont
  {Nauenberg}}\ and\ \bibinfo {author} {\bibfnamefont {D.~J.}\ \bibnamefont
  {Scalapino}},\ }\bibfield  {title} {\bibinfo {title} {Singularities and
  scaling functions at the potts-model multicritical point},\ }\href
  {https://doi.org/10.1103/PhysRevLett.44.837} {\bibfield  {journal} {\bibinfo
  {journal} {Phys. Rev. Lett.}\ }\textbf {\bibinfo {volume} {44}},\ \bibinfo
  {pages} {837} (\bibinfo {year} {1980})}\BibitemShut {NoStop}%
\bibitem [{\citenamefont {Cardy}\ \emph {et~al.}(1980)\citenamefont {Cardy},
  \citenamefont {Nauenberg},\ and\ \citenamefont {Scalapino}}]{Cardy_Potts}%
  \BibitemOpen
  \bibfield  {author} {\bibinfo {author} {\bibfnamefont {J.~L.}\ \bibnamefont
  {Cardy}}, \bibinfo {author} {\bibfnamefont {M.}~\bibnamefont {Nauenberg}},\
  and\ \bibinfo {author} {\bibfnamefont {D.~J.}\ \bibnamefont {Scalapino}},\
  }\bibfield  {title} {\bibinfo {title} {Scaling theory of the potts-model
  multicritical point},\ }\href {https://doi.org/10.1103/PhysRevB.22.2560}
  {\bibfield  {journal} {\bibinfo  {journal} {Phys. Rev. B}\ }\textbf {\bibinfo
  {volume} {22}},\ \bibinfo {pages} {2560} (\bibinfo {year}
  {1980})}\BibitemShut {NoStop}%
\bibitem [{\citenamefont {Salas}\ and\ \citenamefont
  {Sokal}(1997)}]{Salas_Potts}%
  \BibitemOpen
  \bibfield  {author} {\bibinfo {author} {\bibfnamefont {J.}~\bibnamefont
  {Salas}}\ and\ \bibinfo {author} {\bibfnamefont {A.~D.}\ \bibnamefont
  {Sokal}},\ }\bibfield  {title} {\bibinfo {title} {Logarithmic corrections and
  finite-size scaling in the two-dimensional 4-state potts model},\ }\href
  {https://doi.org/10.1023/B:JOSS.0000015164.98296.85} {\bibfield  {journal}
  {\bibinfo  {journal} {Journal of Statistical Physics}\ }\textbf {\bibinfo
  {volume} {88}},\ \bibinfo {pages} {567} (\bibinfo {year} {1997})}\BibitemShut
  {NoStop}%
\bibitem [{\citenamefont {Capponi}\ \emph {et~al.}(2014)\citenamefont
  {Capponi}, \citenamefont {Jahromi}, \citenamefont {Alet},\ and\ \citenamefont
  {Schmidt}}]{Capponi-BW-Tfield}%
  \BibitemOpen
  \bibfield  {author} {\bibinfo {author} {\bibfnamefont {S.}~\bibnamefont
  {Capponi}}, \bibinfo {author} {\bibfnamefont {S.~S.}\ \bibnamefont
  {Jahromi}}, \bibinfo {author} {\bibfnamefont {F.}~\bibnamefont {Alet}},\ and\
  \bibinfo {author} {\bibfnamefont {K.~P.}\ \bibnamefont {Schmidt}},\
  }\bibfield  {title} {\bibinfo {title} {Baxter-wu model in a transverse
  magnetic field},\ }\href {https://doi.org/10.1103/PhysRevE.89.062136}
  {\bibfield  {journal} {\bibinfo  {journal} {Phys. Rev. E}\ }\textbf {\bibinfo
  {volume} {89}},\ \bibinfo {pages} {062136} (\bibinfo {year}
  {2014})}\BibitemShut {NoStop}%
\bibitem [{\citenamefont {Vasilopoulos}\ \emph {et~al.}(2022)\citenamefont
  {Vasilopoulos}, \citenamefont {Fytas}, \citenamefont {Vatansever},
  \citenamefont {Malakis},\ and\ \citenamefont {Weigel}}]{Weigel-diluteBW}%
  \BibitemOpen
  \bibfield  {author} {\bibinfo {author} {\bibfnamefont {A.}~\bibnamefont
  {Vasilopoulos}}, \bibinfo {author} {\bibfnamefont {N.~G.}\ \bibnamefont
  {Fytas}}, \bibinfo {author} {\bibfnamefont {E.}~\bibnamefont {Vatansever}},
  \bibinfo {author} {\bibfnamefont {A.}~\bibnamefont {Malakis}},\ and\ \bibinfo
  {author} {\bibfnamefont {M.}~\bibnamefont {Weigel}},\ }\bibfield  {title}
  {\bibinfo {title} {Universality in the two-dimensional dilute baxter-wu
  model},\ }\href {https://doi.org/10.1103/PhysRevE.105.054143} {\bibfield
  {journal} {\bibinfo  {journal} {Phys. Rev. E}\ }\textbf {\bibinfo {volume}
  {105}},\ \bibinfo {pages} {054143} (\bibinfo {year} {2022})}\BibitemShut
  {NoStop}%
\bibitem [{\citenamefont {Jacobsen}\ and\ \citenamefont
  {Wiese}(2024)}]{PhysRevLett.133.077101}%
  \BibitemOpen
  \bibfield  {author} {\bibinfo {author} {\bibfnamefont {J.~L.}\ \bibnamefont
  {Jacobsen}}\ and\ \bibinfo {author} {\bibfnamefont {K.~J.}\ \bibnamefont
  {Wiese}},\ }\bibfield  {title} {\bibinfo {title} {Lattice realization of
  complex conformal field theories: Two-dimensional potts model with $q>4$
  states},\ }\href {https://doi.org/10.1103/PhysRevLett.133.077101} {\bibfield
  {journal} {\bibinfo  {journal} {Phys. Rev. Lett.}\ }\textbf {\bibinfo
  {volume} {133}},\ \bibinfo {pages} {077101} (\bibinfo {year}
  {2024})}\BibitemShut {NoStop}%
\bibitem [{\citenamefont {Ling}\ \emph {et~al.}(2026)\citenamefont {Ling},
  \citenamefont {Wang}, \citenamefont {Guo},\ and\ \citenamefont
  {Liu}}]{liu2026zenodo}%
  \BibitemOpen
  \bibfield  {author} {\bibinfo {author} {\bibfnamefont {Y.}~\bibnamefont
  {Ling}}, \bibinfo {author} {\bibfnamefont {Y.}~\bibnamefont {Wang}}, \bibinfo
  {author} {\bibfnamefont {W.}~\bibnamefont {Guo}},\ and\ \bibinfo {author}
  {\bibfnamefont {Y.}~\bibnamefont {Liu}},\ }\href
  {https://doi.org/10.5281/zenodo.19034084} {\bibinfo {title} {Data for
  "probing the critical behavior of a sign-problematic model with monte carlo
  simulations"}} (\bibinfo {year} {2026})\BibitemShut {NoStop}%
\end{thebibliography}%
\end{document}